\documentclass[prb,twocolumn,showpacs,amsmath,amssymb,superscriptaddress,floatfix]{revtex4-2}
\usepackage{bm}
\usepackage[pdftex]{graphicx,hyperref}
\hypersetup{colorlinks = true, urlcolor = blue, linkcolor = blue, citecolor = blue}
\usepackage{mathtools}
\usepackage{color}
\usepackage{multirow}
\usepackage{makecell}
\usepackage{titlesec}
\usepackage[normalem]{ulem}

\begin{document}

\title{Pseudospin-Triplet Pairing in Iron-Chalcogenide Superconductors}

\author{Meng Zeng}
\affiliation{Department of Physics, University of California, San Diego, California 92093, USA}

\author{Dong-Hui Xu}
\affiliation{Department of Physics and Chongqing Key Laboratory for Strongly Coupled Physics, Chongqing University, Chongqing 400044, People's Republic of China}
\affiliation{Center of Quantum Materials and Devices, Chongqing University, Chongqing 400044, People's Republic of China}

\author{Zi-Ming Wang}
\affiliation{Department of Physics and Chongqing Key Laboratory for Strongly Coupled Physics, Chongqing University, Chongqing 400044, People's Republic of China}
\affiliation{Center of Quantum Materials and Devices, Chongqing University, Chongqing 400044, People's Republic of China}

\author{Lun-Hui Hu}
\thanks{hu.lunhui.zju@gmail.com}
\affiliation{Department of Physics, The Pennsylvania State University, University Park, Pennsylvania 16802, USA}
\affiliation{Department of Physics and Astronomy, University of Tennessee, Knoxville, Tennessee 37996, USA}

\author{Fu-Chun Zhang}
\affiliation{Kavli Institute for Theoretical Sciences, University of Chinese Academy of Sciences, Beijing 100190, China}
\affiliation{CAS Center for Excellence in Topological Quantum Computation, University of Chinese Academy of Sciences, Beijing 100190, China}

\begin{abstract}	
\begin{center}
\vspace{5pt}
{\bf Abstract}
\end{center}
\vspace{-5pt}
We study superconductivity of electron systems with both spin and pseudospin-1/2 degrees of freedom. By solving linearized gap equations, we derive a weak coupling criterion for the even-parity spin-singlet pseudospin-triplet pairing. It can generally mix with the on-site $s$-wave pairing since both of them belong to the same symmetry representation ($A_{1g}$) and their mixture could naturally give rise to anisotropic intra-band pairing gap functions with or without nodes. This may directly explain why some of the iron-chalcogenide superconductors are fully gapped (e.g.~FeSe thin film) and some have nodes (e.g.~LaFePO and LiFeP).
We also find that the anisotropy of gap functions can be enhanced when the principal rotation symmetry is spontaneously broken in the normal state such as nematicity, and the energetic stabilization of pseudospin-triplet pairings indicates the coexistence of nematicity and superconductivity.
This could be potentially applied to bulk FeSe, where gap anisotropy has been experimentally observed.
\end{abstract}

\maketitle

\noindent
{\bf Introduction}\\
The symmetry principle is one of the most powerful tools to diagnose low-energy electronic band structures, lattice vibrations, and linear responses~\cite{dresselhaus2007group}, and is also valuable to explore various symmetry-breaking ordered phases such as magnetism, charge/spin density-wave, nematicity and superconductivity~\cite{chaikin1995principles}. The crystal symmetry of a solid-state system dictates the normal band structures it hosts near the Fermi level, which could in turn determine the most favorable superconducting pairing symmetry~\cite{sigrist_rmp_1991,frigeri_prl_2004}. This symmetry principle for superconductors (SC) is recently extended to investigate multi-band unconventional superconductivity~\cite{fischer2013gap,ramires_prb_2018,andersen_sa_2020}. 
Interestingly, 
the orbital-independent and orbital-dependent pairings that belong to the same symmetry representation may coexist with each other~\cite{mackenzie_rmp_2003}. Such orbital-dependent pairings have been studied in a wide variety of systems with multi-band character, including
Sr$_2$RuO$_4$~\cite{agterberg_prl_1997}, iron-chalcogenide SCs~\cite{dai_prl_2008,ong_prl_2013,sprau_sci_2017,nica_npjqm_2021}, Cu-doped Bi$_2$Se$_3$~\cite{fu_prl_2010} and half-Heusler compounds~\cite{brydon_prl_2016,yang_prl_2016,savary_prb_2017,yu_prb_2018}, from which the guiding principle by symmetry has been shown to be crucial to understanding the nature of unconventional superconductivity.

A few specific systems can be effectively characterized by a general normal-state model Hamiltonian that contains both spin ($\{\uparrow,\downarrow\}$) and pseudospin ($\{1,2\}$) degrees of freedom, where pseudospin could originate from two atomic orbitals, two sublattices, two layers, or two valleys~\cite{ramires_prb_2018}. 
We start from a spin-singlet centrosymmetric SC to explore the existence of even-parity pseudospin-triplet pairings, for example, $c_{1,\uparrow}(\mathbf{k}) c_{2,\downarrow}(-\mathbf{k}) + c_{2,\uparrow}(\mathbf{k}) c_{1,\downarrow}(-\mathbf{k})  - c_{1,\downarrow}(\mathbf{k}) c_{2,\uparrow}(-\mathbf{k}) - c_{2,\downarrow}(\mathbf{k}) c_{1,\uparrow}(-\mathbf{k})$, and further investigate their valuable roles in tailoring anisotropic pairing gap functions with or without nodes~\cite{scalapino_rmp_2012}. Different from spin-triplet pairings, spin-singlet pseudospin-triplet pairings have not been much explored in real materials since such pairings are usually considered to be energetically unfavorable. 
This is partly due to the common belief that the double degeneracy of the two orbitals is lifted by orbital hybridization so that the orbital-dependent pairing would be severely suppressed under crystal field splitting or electron-electron repulsive interaction. One aim of this work is concerned with the possible condition for the existence of even-parity spin-singlet orbital-dependent pairings, and possible applications to real materials.

On the other hand, the effects of symmetry breaking in unconventional SCs is an important topic that has attracted tremendous interest. The symmetry could be broken explicitly by external fields or strain, or be broken spontaneously from many-body interactions.
Two typical examples are rotational symmetry breaking \cite{fradkin_arcmp_2010,fernandes_arcmp_2019} and time-reversal-symmetry (TRS) breaking~\cite{Sigrist1998,lee_prl_2009,hu_prr_2020,lado_prr_2019,hu_arxiv_2021}. Besides, the interplay between nematicity and superconductivity is yet to be fully understood in some real materials, such as FeSe~\cite{mcqueen_prl_2009,sprau_science_2017}, where gap functions can be highly anisotropic. 
These systems are all multi-band SCs, while symmetry-reducing signatures are experimentally observed above the superconducting transition temperature, which is mainly caused by both crystal field splittings and interaction-induced order parameters (e.g.~nematicity). Thus, discovering the coexistence of nematicity and superconductivity in these multi-band systems can shed new light on understanding the underlying favorable pairing symmetries.

The main finding of this work is that the anisotropic gap functions with or without nodes could be attributed to the mixing of isotropic $s$-wave pairing and even-parity spin-singlet pseudospin-triplet pairing, even though both of them belong to the $A_{1g}$ symmetry representation. For technical conveniences, we adopt an orbital $\mathbf{d}_o(\mathbf{k})$-vector notations~\cite{ong_prl_2013}  to describe the pairing matrix and similarly a $\mathbf{g}_o(\mathbf{k})$-vector for orbital hybridization in the two-orbital subspace ($\{1,2\}$). Solving linearized gap equations, we show that the presence of $\mathbf{g}_o$-vector generally suppresses the superconductivity with orbital $\mathbf{d}_o$-vector except for $\mathbf{d}_o(\mathbf{k})\parallel \mathbf{g}_o(\mathbf{k})$, which is consistent with the concept of superconducting fitness~\cite{ramires_prb_2018}. This sets up weak-coupling criteria for $A_{1g}$-type orbital-dependent pairings that could naturally give rise to anisotropic gap functions in real superconducting materials.
Moreover, we reveal a deep connection between two-orbital nematic SC and pseudospin-triplet pairings. 
Within the mean-field theory for electron-electron repulsive interactions, the nematic order develops in the orbital subspace at $T<T_{\text{nem}}$, which also contributes to the total orbital hybridization, $\mathbf{g}_{\text{tot}} = \mathbf{g}_o+\mathbf{g}_{\text{nem}}$. 
This leads to the stabilization of a nematic orbital $\mathbf{d}_o$-vector for $\mathbf{d}_o(\mathbf{k})\parallel \mathbf{g}_{\text{tot}}(\mathbf{k})$, indicating the coexistence of nematicity and superconductivity. The direct applications to FeSe~\cite{mcqueen_prl_2009,sprau_science_2017} are also discussed.
We also generalize it to a two-valley system with $C_6$ breaking terms (e.g.,~Kekul\'e distortion). In the end, we also predict an orbital-polarized superconducting state.

\begin{table*}[t]
	\begin{ruledtabular}
		\begin{tabular}{c|c|c|c|c|c}
			$C_n$ & $J=-J \, (\text{mod }n)$ & $\Psi_s(\mathbf{k})=\Psi_s(-\mathbf{k})$ & $\mathbf{d}_o^1(\mathbf{k})=\mathbf{d}_o^1(-\mathbf{k})$ & $\mathbf{d}_o^2(\mathbf{k})=-\mathbf{d}_o^2(-\mathbf{k})$ & $\mathbf{d}_o^3(\mathbf{k})=\mathbf{d}_o^3(-\mathbf{k})$ \\ \hline
			\multirow{2}{*}{$n=2$}
			& $J=0$ &  $1,k_x^2,k_y^2,k_z^2,k_xk_y$ & $1,k_x^2,k_y^2,k_z^2,k_xk_y$ & $k_z,k_zk_x^2,k_zk_y^2,k_z^3,k_zk_xk_y$ & $1,k_x^2,k_y^2,k_z^2,k_xk_y$ \\ 
			& $J=1$ &  $k_xk_z,k_yk_z$ & $k_xk_z,k_yk_z$ & $k_x,k_y$ & $k_xk_z,k_yk_z$ \\ \hline  
			{$n=3$} 
			& $J=0$ &  $1,k_x^2+k_y^2,k_z^2$ & $E_g$ representation & $k_z$ & $E_g$ representation \\ \hline
			\multirow{2}{*}{$n=4$} 
			& $J=0$ & $1, k_x^2+k_y^2, k_z^2$ & $k_x^2-k_y^2, k_xk_y$   & $k_z,k_z(k_x^2+k_y^2),k_z^3$ & $k_x^2-k_y^2, k_xk_y$ \\
			& $J=2$ & $k_x^2-k_y^2,k_xk_y$    & $1, k_x^2+k_y^2, k_z^2$ & $k_z(k_x^2-k_y^2),k_zk_xk_y$   & $1, k_x^2+k_y^2, k_z^2$ \\ \hline
			\multirow{2}{*}{$n=6$} 
			& $J=0$ & $1,k_x^2+k_y^2,k_z^2$ & $E_g$ representation & $k_z$ & $E_g$ representation \\
			& $J=3$ & $(k_x+ik_y)^3,(k_x-ik_y)^3$ & $E_g$ representation & $k_x^3-3k_xk_y^2,3k_x^2k_y-k_y^3$ & $E_g$ representation \\ 
		\end{tabular}
		\caption{\label{tab:symmetry-constraint} Classification of spin-singlet pairing potentials for Eq.~\eqref{eq-J-2-pairing}. Here we consider a spin-singlet two-orbital superconductors with $\{d_{xz},d_{yz}\}$-orbitals.
			Based on the $n$-fold rotation symmetry $C_n$ about $z$-axisd and TRS, we have $J = -J$ mod $n$, which leads to all the pairing channels with orbital-independent $\Psi_s(\mathbf{k})$ and orbital-dependent $\mathbf{d}_o(\mathbf{k})$-vector in Eq.~\eqref{eq-J-2-pairing}. Here, for $J=0$ pairing subspace of $C_3$, the $(\mathbf{d}_o^1(\mathbf{k}),\mathbf{d}_o^3(\mathbf{k}))$ forms a two-dimensional $E_g$ representation, where the basis functions are $(k_x^2-k_y^2,k_xk_y)$ and $(k_yk_z,k_xk_z)$.
			For $J=0$ pairing subspace of $C_6$, the $(\mathbf{d}_o^1(\mathbf{k}),\mathbf{d}_o^3(\mathbf{k}))$ forms a two-dimensional $E_g$ representation, where the basis functions are $(k_yk_z,k_xk_z)$; for the $J=3$ pairing subspace of $C_6$, the $(\mathbf{d}_o^1(\mathbf{k}),\mathbf{d}_o^3(\mathbf{k}))$ forms a two-dimensional $E_g$ representation, where the basis functions are $(k_x^2-k_y^2,k_xk_y)$.
		}
	\end{ruledtabular}
\end{table*}

\vspace{10pt}
\noindent
{\bf Results}\\
{\bf Classification of Spin-singlet Orbital-triplet pairings.}
To explore the weak-coupling criterion for the energetically favorable even-parity spin-singlet pseudospin-triplet pairing, we consider the mean-field pairing Hamiltonian,
\begin{align}
\mathcal{H}_{\Delta} = \sum_{\mathbf{k}}\sum_{s_1a,s_2b}
\Delta_{s_1,s_2}^{a,b}(\mathbf{k}) 
F^\dagger_{s_1a,s_2b}(\mathbf{k}) +\text{h.c.},
\end{align}
where $F^\dagger_{s_1a,s_2b}(\mathbf{k})=c^\dagger_{s_1a}(\mathbf{k})  c^\dagger_{s_2b}(-\mathbf{k})$ is the creation operator of Cooper pairs, $s_1, s_2$ are indices for spins and $a,b$ are for pseudospins (e.g.,~two orbitals $\{1,2\}$).
A general pairing potential of a two-band model is a four-by-four matrix~\cite{ramires_prb_2018}.
In particular, the spin-singlet pairing function $\Delta_{s_1,s_2}^{a,b}(\mathbf{k})=f(\mathbf{k})M_{a,b}(\bf k)(i\sigma_2)_{s_1,s_2}$ consists of the angular form factor $f(\mathbf{k})$ and $M_{a,b}(\bf k)$ in the orbital channel. The spin-singlet pairings are not mixed with spin-triplet pairings in the absence of spin-orbit coupling (SOC).
In analogy to spin-triplet SCs, for the technical convenience, we then use an orbital $\mathbf{d}_o(\mathbf{k})$-vector for the spin-singlet orbital-dependent pairing potential \cite{ong_prl_2013},
\begin{align}\label{eq-J-2-pairing}
\hat{\Delta}_{tot}(\mathbf{k})= [\Delta_s\Psi_s(\mathbf{k})\tau_0+\Delta_o(\mathbf{d}_o(\mathbf{k})\cdot\bm{\tau})](i\sigma_2),
\end{align} 
where $\Delta_s$ and $\Delta_o$ are pairing strengths in orbital-independent and orbital-dependent channels, respectively. 
Here $\bm{\tau}$ and $\bm{\sigma}$ are Pauli matrices acting on the orbital and spin subspace, respectively, and $\tau_0$ is a 2-by-2 identity matrix.
When both $\Delta_s$ and  $\Delta_o$ are real,  a real orbital $\mathbf{d}_o(\mathbf{k})$-vector preserves TRS while a complex one spontaneously breaks TRS ($\mathcal{T}=i\tau_0\sigma_2\mathcal{K}$ with $\mathcal{K}$ being complex conjugate).
The Fermi statistics requires $\Psi_s(\mathbf{k}) = \Psi_s(-\mathbf{k})$, $d_o^{1,3}(\mathbf{k})=d_o^{1,3}(-\mathbf{k})$ and $d_o^2(\mathbf{k})=-d_o^2(-\mathbf{k})$. 
In other words, $d_o^2(\mathbf{k})$ describes odd-parity spin-singlet orbital-singlet pairings and the other two are for even-parity spin-singlet orbital-triplet pairings.
Moreover, we provide an alternative definition of orbital $\mathbf{d}_o$-vectors in Appendix A.
Even though the orbital-independent part $\Psi_s({\bf k})$ is also ``orbital-triplet'' by statistics, it is completely trivial. Hereafter, we only refer to $d_o^{1}({\bf k})$ and $d_o^{3}({\bf k})$ as orbital-triplet pairings~\cite{ong_pnas_2016}.

In addition, the basis functions for both $\Psi_s(\mathbf{k})$ and orbital $\mathbf{d}_o(\mathbf{k})$-vectors in Eq.~\eqref{eq-J-2-pairing} could be classified by crystalline symmetry. 

Under the action of an $n$-fold rotation operator $C_n$ about the $z$-axis, the  pairing potential $\hat{\Delta}(\mathbf{k})$ transforms as 
\begin{align} \label{eq-pairing-classification}
{\cal D}[C_n]\,
\hat{\Delta}_J(\mathbf{k})\,
({\cal D}[C_n])^T
=e^{i\frac{2\pi}{n}J}\,
\hat{\Delta}_J(C_n^{-1}\mathbf{k}),
\end{align} 
where ${\cal D}[C_n]$ is the corresponding matrix representation, $J$ is the orbital angular momentum quantum number, and also labels the irreducible representations of the $C_n$ point group.
For example, $J=0$ is for $A$ representation and $J=2$ is for $B$ representation. 
Firstly, the TRS requires the coexistence of $\hat{\Delta}_J$ and $\hat{\Delta}_{-J}$ with equal weight. If the rotation symmetry $C_n$ is further imposed, then $J$ and $-J$ have to be equivalent modulo $n$, i.e. $J\equiv -J \text{ mod }n$. 
The results for the basis functions of $\Psi_s(\mathbf{k})$ and $\mathbf{d}_o(\mathbf{k})$ are summarized in Table~(\ref{tab:symmetry-constraint}) for a two-band SC with the $\{d_{xz},d_{yz}\}$-orbitals.
In this case, ${\cal D}[C_n] = [ \cos(\tfrac{2\pi}{n})\tau_0 - i \sin(\tfrac{2\pi}{n}) \tau_2 ] \otimes \sigma_0$. For instance, ${\cal D}[C_4]=-i\tau_2\otimes\sigma_0$ explains that both $\Delta_o \tau_1$ and $\Delta_o \tau_3$ are d-wave-like pairing states~\cite{agterberg_prl_2017b}.

At the mean-field level, the Bogoliubov de-Gennes (BdG) Hamiltonian is given by
\begin{align}
    \mathcal{H}_{\text{BdG}} = \begin{pmatrix}
      \mathcal{H}_0(\mathbf{k}) & \hat{\Delta}_{tot}({\bf k}) \\
      \hat{\Delta}_{tot}^\dagger({\bf k}) & -\mathcal{H}_0^\ast(-\mathbf{k})
    \end{pmatrix},
\end{align}
where $\mathcal{H}_0(\mathbf{k})$ represents a two-band normal-state Hamiltonian with both spin and pseudospin degrees of freedom. 

In general, the BdG Hamiltonian is also invariant under the $C_n$ rotation symmetry, i.e., ${\cal D}_{\text{BdG}}[C_n] \, {\cal H}_{\text{BdG}}({\bf k}) \, ({\cal D}_{\text{BdG}}[C_n])^\dagger = {\cal H}_{\text{BdG}}(C_n^{-1}{\bf k})$ when we define ${\cal D}_{\text{BdG}}[C_n] = \begin{pmatrix}
  {\cal D}[C_n]  & 0 \\
  0  &  e^{i\tfrac{2\pi}{n}J} ({\cal D}[C_n])^\ast
\end{pmatrix}$ based on Eq.~\eqref{eq-pairing-classification}. 

Here we assume both inversion and time-reversal symmetries are preserved. To be specific, we consider a SOC-free Hamiltonian, 
\begin{equation}
\begin{split}
\mathcal{H}_0(\mathbf{k}) = \epsilon(\mathbf{k})\tau_0\sigma_0 + \lambda_o(\mathbf{g}_o(\mathbf{k})\cdot\bm{\tau})\sigma_0,
\end{split}
\label{Eq::normal_H}
\end{equation}
where the basis is $\psi_{\mathbf{k}}^\dagger=(c_{1,\uparrow}^\dagger(\mathbf{k}),  c_{1,\downarrow}^\dagger(\mathbf{k}),c_{2,\uparrow}^\dagger(\mathbf{k}),c_{2,\downarrow}^\dagger(\mathbf{k}))$,
$\epsilon(\mathbf{k}) = (k_x^2+k_y^2)/2m -\mu$ is the band energy measured relative to the chemical potential $\mu$, $m$ is the effective mass, $\lambda_o$ represents the orbital hybridization and $\mathbf{g}_o(\mathbf{k})=(g_1(\mathbf{k}),g_2(\mathbf{k}),g_3(\mathbf{k}))$. And the $g_3$-component leads to the different effective masses of different orbitals. As mentioned earlier, this vector notation is just for the technical convenience. Besides, the $g_1$ and $g_2$ components are determined by symmetries. For example, TRS requires $g_{1,3}(\mathbf{k})=g_{1,3}(-\mathbf{k})$ and $g_{2}(\mathbf{k})=-g_{2}(-\mathbf{k})$.
If inversion symmetry (IS) is present, $g_{2}(\mathbf{k})$ (or $g_{1}(\mathbf{k})$) must vanish for $\mathcal{I}=\tau_0\sigma_0$ (or $\mathcal{I}=\tau_3\sigma_0$), which is the same as the constraint for the orbital $\mathbf{d}_o$-vector. The more explicit form of ${\bf g}_o({\bf k})$ is determined by other crystal symmetries.

In general, the pseudospin-triplet (i.e.~orbital-triplet) pairing state shares some similarities with the spin-triplet pairing state~\cite{Smidman_ropip_2017}. To show that, we first discuss the superconducting quasi-particle spectrum of orbital-triplet SCs in the absence of band-splitting caused by orbital hybridizations, i.e., ${\bf g}_o({\bf k})=0$ for Eq.~\eqref{Eq::normal_H}. In this case, the superconducting gaps on the Fermi surface are
\begin{align} \label{eq-sc-gap-complex-d}
E({\bf k}) = \pm \vert\Delta_o\vert
\sqrt{\vert\mathbf{d}_o(\mathbf{k})\vert^2 \pm \vert \mathbf{d}_o^\ast(\mathbf{k}) \times \mathbf{d}_o(\mathbf{k}) \vert },
\end{align}
for the $\Delta_s=0$ limit. This indicates that there are two distinct gaps if TRS is spontaneously broken. In the following, we mainly focus on the time-reversal-invariant superconducting states, i.e., real ${\bf d}_o$-vectors, for which the classification of pairing potentials is shown in Table~(\ref{tab:symmetry-constraint}) based on Eq.~\eqref{eq-pairing-classification}. We will show the interplay between $\Delta_s$ and $\Delta_o$ can lead to anisotropic superconducting gaps on different Fermi surfaces.
Moreover, its stability against orbital-hybridization, electron-electron interactions, and applications to real materials will be discussed in detail as follows. We will also briefly comment on the effects of TRS-breaking in the end.

\vspace{10pt}
\noindent
{\bf Stability for spin-singlet orbital-triplet pairings.}
We apply the weak-coupling scheme~\cite{ramires_prb_2018} for spin-singlet orbital-triplet pairings against crystal field splittings, which cause orbital hybridizations [i.e.~the ${\bf g}_o({\bf k})$ term in Eq.~\eqref{Eq::normal_H}]. We analytically calculate the superconductivity instability for the orbital $\mathbf{d}_o$-vector by BCS decoupling scheme. 
The superconducting transition temperature $T_c$ of orbital-dependent pairing channels is calculated by solving the linearized gap equation,
\begin{align}
\begin{split}
\Delta_{s_1,s_2}^{a,b}(\mathbf{k}) &= -\frac{1}{\beta} \sum_{\omega_n}\sum_{s_1'a',s_2'b'} V^{s_1a,s_2b}_{s_1'a',s_2'b'}(\mathbf{k},\mathbf{k}') \times \\
&\left\lbrack G_e(\mathbf{k}',i\omega_n) \hat{\Delta}(\mathbf{k}') G_h(-\mathbf{k}',i\omega_n) \right\rbrack_{s_1'a',s_2'b'},
\end{split}
\end{align}
where $\beta=1/k_BT$, $G_e(\mathbf{k},i\omega_n)=[i\omega_n-\mathcal{H}_0(\mathbf{k})]^{-1}$ is the Matsubara Green's function for electrons with $\omega_n=(2n+1)\pi/\beta$ and $G_h(\mathbf{k},i\omega_n)=-G_e^\ast(\mathbf{k},i\omega_n)$. 
We expand the attractive interactions as $V^{s_1a,s_2b}_{s_1'a',s_2'b'}(\mathbf{k},\mathbf{k}') = -\sum_{\Gamma,l} \lbrack v^{\Gamma}\mathbf{d}_o^{\Gamma,l}(\mathbf{k})\cdot\bm{\tau}i\sigma_2 \rbrack_{s_1a,s_2b} \lbrack \mathbf{d}_o^{\Gamma,l}(\mathbf{k}')\cdot\bm{\tau}i\sigma_2 \rbrack_{s_1'a',s_2'b'}$ with $v^\Gamma>0$.
Here $\Gamma$ labels the irreducible representation with $l=1,2,...,\mathrm{Dim}\ \Gamma$. In this work we focus on 1d representations, i.e. Dim $\Gamma=1$, which already include many interesting cases and are sufficient for the applications discussed in later sections. Due to the possible existence of multiple pairing channels belonging to different representations, each channel has its own critical temperature $T_c^\Gamma$, the largest of which becomes the actual critical temperature of the system. In the weak-coupling theory, $T_c^\Gamma$ follows the standard BCS form and is solely determined by the corresponding pairing interaction $v^\Gamma$ in that particular channel.
To the leading order of $\lambda_ok_F^2/\mu$ ($k_F=\sqrt{2m\mu}$), the equation for $T_c$ for the channel $\Gamma$ reads (see details in the Methods section), 
\begin{align}\label{eq-tc-lambdao}
\ln\left(\frac{T_c^{\Gamma}}{T_{c0}} \right) = \int_{S} d\Omega \, \mathcal{C}_0(T_c) \left( \left\vert {\mathbf{d}}_o^{\Gamma} \right\vert^2 -  \left\vert {\mathbf{d}}_o^{\Gamma}\cdot\hat{\mathbf{g}}_o \right\vert^2 \right),
\end{align}
where $T_{c0}$ is the critical temperature for $\lambda_o=0$, $\Omega$ is the solid angle of $\mathbf{k}$, $\hat{\mathbf{g}}_o=\mathbf{g}_o(\mathbf{k})/\vert\mathbf{g}_o(\mathbf{k})\vert$ are normalized vectors.
Here we take $\int_{S} d\Omega \, \vert {\mathbf{d}}_o^{\Gamma} \vert^2 =1$.
And $\mathcal{C}_0(T_c)=\text{Re}\lbrack \psi^{(0)}(\tfrac{1}{2}) - \psi^{(0)}(\tfrac{1}{2}+i\tfrac{\lambda_o\vert\mathbf{g}(\mathbf{k})\vert}{2\pi k_BT_c}) \rbrack$, where $\psi^{(0)}(z)$ is the digamma function.

\begin{figure}[t]
	\centering
	\includegraphics[width=0.9\linewidth]{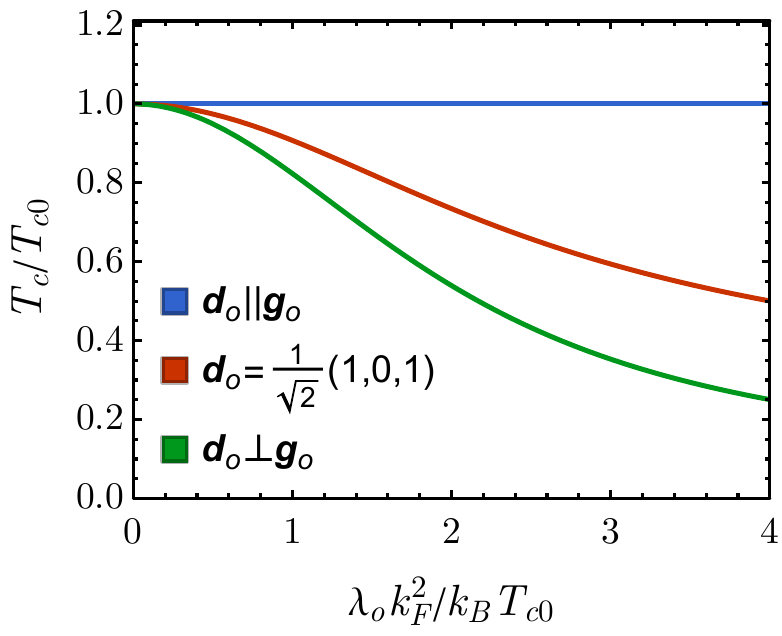}
	\caption{Stability of orbital $\mathbf{d}_o$-vectors vs orbital hybridization $\lambda_0 $ in Eq.~\eqref{Eq::normal_H}. Shown are the transition temperature $T_c/T_{c0}$ as a function of $\lambda_ok_F^2/k_BT_{c0}$ for $\mathbf{g}_o(\mathbf{k})=(2k_xk_y,0,k_x^2-k_y^2)$. $T_{c0}$ is $T_c$ at $\lambda_0=0$. 
		The curves from top to bottom correspond to $\mathbf{d}_o(\mathbf{k})=k_F^{-2} (2k_xk_y,0,k_x^2-k_y^2)$, $\mathbf{d}_o(\mathbf{k})=\frac{1}{\sqrt{2}}(1,0,1)$, and $\mathbf{d}_o(\mathbf{k})=k_F^{-2}(k_x^2-k_y^2,0,-2k_xk_y)$, respectively.
	}
	\label{fig2}
\end{figure}

We now discuss its implications. In general, the $\lambda_o$-term describes a pair-breaking term, since $\mathcal{C}_0(T_c)\le 0$ and it monotonically decreases as $\lambda_o$ increases, hence the right-hand side of Eq.~\eqref{eq-tc-lambdao} suppresses $T_c$ in general. However, if we focus on one-dimensional representations, i.e. Dim $\Gamma=1$, it is straightforward to see that $\mathbf{d}_o^\Gamma \parallel \mathbf{g}_o$ can lead to $T_c=T_{c0}$ for any value of $\lambda_o$, which indicates that the orbital  $\mathbf{d}_o$-vector that is parallel with $\mathbf{g}_o$ is unaffected by the orbital hybridizations. It is worth mentioning that due to the possible suppression of $T_c$, depending on the relation between $\mathbf{d}_o^\Gamma$ and $\mathbf{g}_o$ the leading instability channel at $\lambda_o=0$ could be suppressed more than some of the other coexisting channels and may eventually become sub-leading. This interesting behavior is discussed further in Appendix C. For notional simplicity, we will drop the representation index $\Gamma$ when there is no danger of confusion.
Choosing $\mathbf{g}_o(\mathbf{k})=(2k_xk_y,0,k_x^2-k_y^2)$, the numerical results are shown in Fig.~\ref{fig2}.
The black line confirms that $T_c$ is unaffected as $\lambda_ok_F^2/k_BT_{c0}$ increases for $\mathbf{d}_o(\mathbf{k})=k_F^{-2}(2k_xk_y,0,k_x^2-k_y^2)$, which is the unconventional $A_{1g}$ pairing.
However, $T_c$ for other $\mathbf{d}_o$-vectors are severely suppressed.
The light-blue line is for $\mathbf{d}_o(\mathbf{k})=\frac{1}{\sqrt{2}}(1,0,1)$, and the light-orange line for $\mathbf{d}_o(\mathbf{k})=k_F^{-2}(k_x^2-k_y^2,0,-2k_xk_y)$.
Therefore, we conclude that the orbital $\mathbf{d}_o$-vector could exist in SCs with two active orbitals that are not fully degenerate.
This is similar to spin-triplet SCs, where the $A_{1g}$-type spin $\mathbf{d}_s$-vector could exist in noncentrosymmetric SCs because $\mathbf{d}_s\parallel \mathbf{g}_s$ is optimally satisfied \cite{frigeri_prl_2004,ramires_prb_2018}. 

It is worth mentioning that the results presented above is using a continuum form of the Hamiltonian based on $\mathbf{k}\cdot\mathbf{p}$ theory.  For real materials, given the interaction on the lattice, the components of the interaction in terms of the basis functions of the representations might not be exactly the same with the form of the vector $\mathbf{g}_o$. As a result, the parallel condition presented above may not be exactly satisfied. However, the theory developed in this work is generally applicable and the extend to which the parallel condition holds can still be a useful criterion for the most favorable pairing.

Next, we include $\Delta_s$, and investigate the coupling between $\Psi_s$ and $\mathbf{d}_o$.
Solving the coupled linearized gap equations up to $(\lambda_ok_F^2/\mu)^2$ order (see details in Appendix C, we find that the results from Eq.~\eqref{eq-tc-lambdao} are still correct. 
Besides, the magnitude of orbital $\mathbf{d}_o$-vectors might be determined as $ \mathbf{d}_o(\mathbf{k}) = \Psi_s(\mathbf{k}) \hat{\mathbf{g}}_o(\mathbf{k}) $.
It implies that $\Psi_s$ and $\mathbf{d}_o$ belong to the same representation of crystalline groups. Therefore, the stability of orbital ${\bf d}_o$-vector by Eq.~\eqref{eq-tc-lambdao} indicates the symmetry principle for spin-singlet orbital-triplet pairings. 

We now explain Eq.~\eqref{eq-tc-lambdao} from the band picture.
Within the band basis, the pairing potential in the orbital subspace becomes 
$\hat{\Delta}_{\text{band}}(\mathbf{k})  =  U^\dagger(\mathbf{k})  \left\lbrack \Delta_s\Psi_s(\mathbf{k})\tau_0 +  \Delta_o(\mathbf{d}_o(\mathbf{k})\cdot\bm{\tau}) \right\rbrack U(\mathbf{k})$, 
where $U(\mathbf{k})$ is the unitary matrix in the orbital subspace,
$U^\dagger(\mathbf{k}) \lbrack  \epsilon(\mathbf{k})\tau_0 + \lambda_o(\mathbf{g}_o(\mathbf{k})\cdot\bm{\tau})\rbrack U(\mathbf{k}) = \text{Diag}[ E_+(\mathbf{k}),  E_-(\mathbf{k})]$, with the normal band dispersion 
\begin{equation}
    E_\pm(\mathbf{k}) = \epsilon(\mathbf{k}) \pm \lambda_o\vert \mathbf{g}_o(\mathbf{k})\vert.
\end{equation}
The intra-orbital pairing naturally gives rise to the intra-band pairing. However, it is different for orbital-dependent pairings. To show that, we decompose the orbital $\mathbf{d}_o$-vector, $\mathbf{d}_o(\mathbf{k}) = d_{\parallel}(\mathbf{k})  \hat{\mathbf{g}}_o(\mathbf{k}) +\mathbf{d}_{\perp}(\mathbf{k})$,
where $d_{\parallel}(\mathbf{k})= \mathbf{d}_o(\mathbf{k}) \cdot \hat{\mathbf{g}}_o(\mathbf{k})  $ and $\mathbf{d}_{\perp}(\mathbf{k})\cdot \hat{\mathbf{g}}_o(\mathbf{k})  = 0$. 
We find that the $d_{\parallel}$-part gives rise to the intra-band pairing, while the $\mathbf{d}_{\perp}$-part leads to the inter-band pairing (see Appendix D).
If the band splitting is much larger than the pairing gap ($\lambda_o k_F^2 \gg \Delta_o$), the inter-band pairing is not energetically favorable in the weak-coupling pairing limit. 
It means that the inter-band pairing will be severely suppressed if we increase the orbital hybridization $\lambda_o$, consistent with Eq.~\eqref{eq-tc-lambdao} and results in Fig.~\ref{fig2}. Now if we again include the orbital-independent pairing part $\Delta_s\Psi_s(\mathbf{k}\tau_0 i\sigma_2)$, the relation between $\mathbf{d}_o$ and $\Psi_s(\mathbf{k})$ obtained previously from solving the coupled linearized gap equation (see Appendix C) can also be reproduced in the band picture by considering the maximization of the condensation energy. The total condensation energy per volume and per spin of the two intra-band pairings is given by
\begin{align}
\begin{split}
\delta E  &=  N_+\sum_{\mathbf{k}\in \text{FS}_+} \left(\Delta_s\Psi_s(\mathbf{k}) +\Delta_od_{\parallel}(\mathbf{k}) \right)^2 \\
&+N_-\sum_{\mathbf{k}\in \text{FS}_-}\left(\Delta_s\Psi_s(\mathbf{k}) -\Delta_od_{\parallel}(\mathbf{k}) \right)^2,
\end{split}
\end{align}
where $N_\pm$ are the density of states on the two Fermi surfaces ($E_{\pm}$). And $\Delta_s\Psi_s(\mathbf{k}) \pm \Delta_od_{\parallel}(\mathbf{k})$ are the pairing gaps on these two Fermi surfaces. In order to maximize $\delta E$, we have $d_{\parallel}(\mathbf{k})=\text{sign}[(N_+-N_-)\Delta_s\Delta_o]\Psi_s(\mathbf{k})$ (See Appendix D for details). Even though the intra-orbital pairing and the orbital-triplet pairing belong to the same symmetry representation, the different ${\bf k}$-dependencies of $\Psi_s(\mathbf{k})$ and $d_{\parallel}(\mathbf{k})$ can naturally lead to the anisotropic superconducting gap on the Fermi surface observed in experiments.

\begin{figure*}[t]
	\centering
	\includegraphics[width=0.95\linewidth]{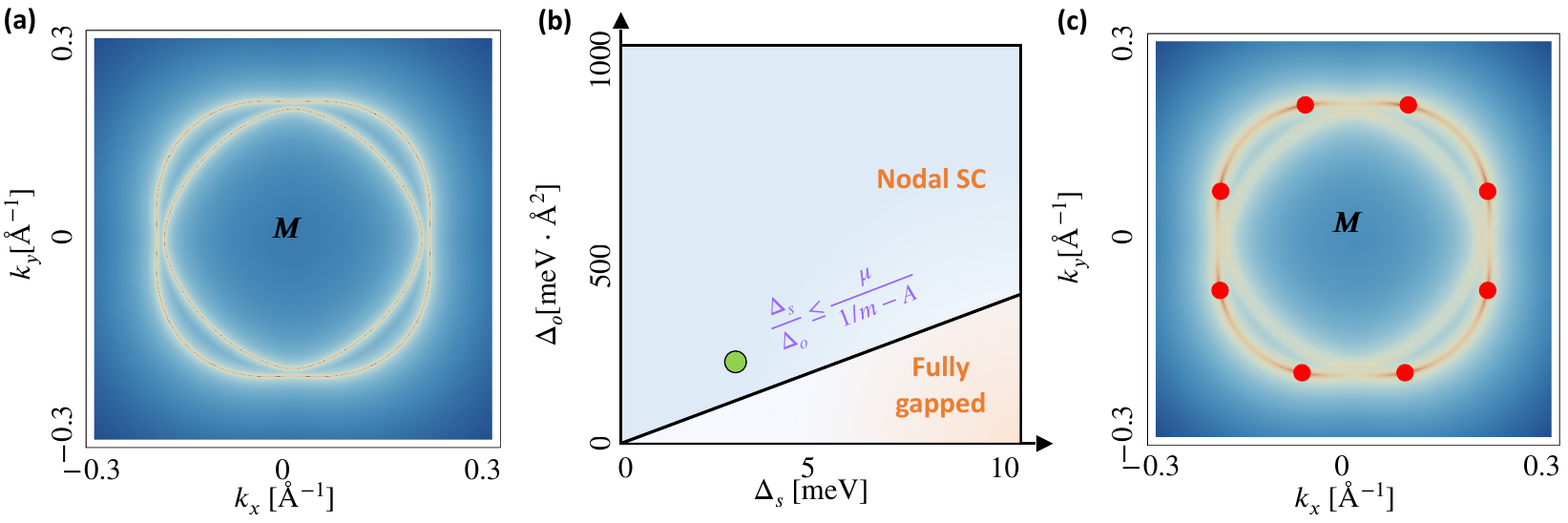}
	\caption{The application to iron-chalcogenide superconductors with/without linear Dirac nodes. In (a), the two-electron pockets around the $M$ point. For zero spin-orbit coupling, $v_{\text{so}}=0$, (b) shows the phase diagram as a function of the intra-orbital pairing $\Delta_s$ and the inter-orbital pairing $\Delta_o$. For the gap parameters represented by the green dot in (b), the nodal superconductor is exhibited in (c), where the eight dark red points represent the chiral symmetry-protected Dirac nodes. 
	}
	\label{fig5}
\end{figure*}


\vspace{10pt}
\noindent
{\bf Applications to superconductors with/without nodes.}
As a consequence of the mixing of the orbital-independent pairing ($\Delta_s$) and orbital-dependent pairing ($\Delta_o$) discussed in the previous section, there could be a nodal SC. In this section, we apply the results of the previous section to study superconductors with two orbitals, where $\Delta_s$ and $\Delta_o$ coexist. It is shown that the anisotropic gap functions with/without nodes depend on the ratio of $\Delta_s$ and $\Delta_o$ superconducting order parameters. Our weak-coupling theory might have potential applications to some of the nodal/nodeless SCs in the iron-chalcogenides family.
For example, the angle-resolved photoemission spectroscopy (ARPES) measurements indicate a nontrivial superconducting gap anisotropy for the monolayer FeSe thin film~\cite{zhang_prl_2016}. The penetration depth measurements on both LaFePO~\cite{fletcher_prl_2009} and LiFeP~\cite{hashimoto_prl_2012} show a linear dependence on $T$, suggesting the presence of superconducting gap nodes.

As an example, we consider the pairing potential in Eq.~\eqref{eq-J-2-pairing} for monolayer FeSe, where there is no hole pocket around the $\Gamma$-point, and a two-spin two-orbital model has been shown to be a good approximation around the electron pockets near the $M$ point of the Brillouin zone (two Fe unit cell).
The density functional theory calculations show that there are four bands around the $M$ point, giving rise to only two electron pockets.
In the one Fe unit cell, there is one pocket near the $X$ and $Y$ points, respectively. After folding with respect to the unit cell with two Fe, we obtain two pockets around the $M$ point. Considering spin degrees of freedom, it naturally resembles a $C_{4z}$-invariant two-orbital model~\cite{nakayama_prb_2018},
\begin{align}
\begin{split}
\mathcal{H}_{M}({\bf k}) &=  \lbrack \epsilon({\bf k}) \tau_0 + A k_x k_y \tau_z \rbrack \sigma_0 \\
&+ v_{\text{so}} \tau_x\lbrack k_x \sigma_y + k_y\sigma_x \rbrack,
\end{split}
\end{align}
where $\epsilon({\bf k})= (k_x^2+k_y^2)/(2m)-\mu$ with $m>0$ the effective mass, $A$ leads to the anisotropic effective mass (i.e.,~orbital hybridization), and $v_{\text{so}}$ represents SOC that still preserves inversion symmetry.
These four states are degenerate at the $M$ point since they form the four-dimensional representation of the space group No.~129 ($P4/nmm$)~\cite{eugenio_prb_2018}.
We take the parameters for the FeSe thin film as $\mu=55$ meV, $1/(2m)=1375$ meV$\cdot$\AA$^2$, $A=600$ meV$\cdot$\AA$^2$ and $v_{\text{so}}\le15$ meV$\cdot$\AA~\cite{nakayama_prb_2018}. The SOC is very weak to open a tiny gap along the $k_x=0$ and $k_y=0$ lines, shown in Fig.~\ref{fig5} (a). As what we expect, it shows two $C_{4z}$ rotational-invariant Fermi surfaces, and the maximal gap, which is induced by the $z$-component of the $\mathbf{g}_o$ vector, is around $12$ meV along the (11) and (1$\bar{1}$) directions. This is larger than the typical superconducting gaps in iron-chalcogenide SCs ($\sim$ 4 meV), implying that the effect of the orbital hybridization on the pairing symmetries should not be neglected.

We now use the criteria derived above (Eq.~\eqref{eq-tc-lambdao}) to examine the superconducting states. Specifically, the weak-coupling criterion indicates that the most favorable pairing to characterize the anisotropic superconducting gap is the $A_{1g}$-type $s$-wave pairing symmetry,
\begin{align}
\hat{\Delta}(\mathbf{k})= [\Delta_s\tau_0+\Delta_o k_x k_y\tau_z](i\sigma_2). 
\end{align}
The ratio between $\Delta_s$ and $\Delta_o$ determines the superconducting nodal structure. To simplify the analysis, we turn off the weak SOC. In the band basis, the dispersion of $\mathcal{H}_{M}({\bf k})$ is $\epsilon_\pm ({\bf k}) = (k_x^2+k_y^2)/(2m) \pm A \vert k_xk_y\vert -\mu$. Here $\pm$ label the band index. Projecting $\hat{\Delta}(\mathbf{k})$ onto the bands leads to $\Delta_\pm = \Delta_s \pm \Delta_o \vert k_x k_y\vert$. Given that $\Delta_s,\Delta_o>0$, nodal points can only appear for $\Delta_{-}$ on the ``$-$'' band. The nodal condition would be $|k_xk_y|=\Delta_s/\Delta_o$ has solution on the FS given by $\epsilon_-(\bf k)=0$. By using the mathematical inequality $k_x^2+k_y^2\leq 2|k_xk_y|$, it can be shown that the nodal condition is given by,
\begin{align}
 \frac{\Delta_s}{\Delta_o} \leq \frac{\mu}{1/m-A},
\end{align}
which is shown in Fig.~\ref{fig5} (b). 
In general, the ratio $\Delta_s/\Delta_o$ should depend on both interaction strength in each pairing channel and the orbital hybridization strength.
This gives rise to the condition of nodal $A_{1g}$-type $s$-wave superconducting states. Therefore, it could not only explain the anisotropic gap functions observed in the FeSe thin film (fully gapped) but also the nodal superconductivity in LaFePO and LiFeP. 
Around one linear Dirac node, the effective Hamiltonian up to linear-$k$ can be mapped out as 
\begin{align}
\mathcal{H}_{D} = k_1 \tilde{\sigma}_0\tilde{\tau}_z + k_2\tilde{\sigma}_y\tilde{\tau}_y,
\end{align}
where $k_1,k_2$ are liner combinations of $k_x$ and $k_y$. All the other Dirac nodes are related to this one by reflection symmetries. Then, we only need to focus on $\mathcal{H}_{D}$, which is a Dirac Hamiltonian with topological charge (winding number) $\pm 2$, whose node is protected by the chiral symmetry (i.e., the product of time-reversal symmetry and particle-hole symmetry). The $2\mathbb{Z}$ winding number is due to the presence of inversion symmetry and time-reversal symmetry. 
To analytically show the topology of Dirac nodes, we apply perturbation analysis with respect to PT symmetry (i.e.,~the product of time-reversal symmetry and inversion symmetry) and Chiral symmetry. Note that the PT symmetry can be also $C_{2z}T$ symmetry for a 2D or quasi-2D SC. The projected symmetry representations are given by
$ \text{PT} = \tilde{\sigma}_y \tilde{\tau}_0$ and $\mathcal{C} = \tilde{\sigma}_y \tilde{\tau}_x$.
As expected, the PT symmetry commutes with $\mathcal{H}_{D}$, while the Chiral symmetry anti-commutes with $\mathcal{H}_{D}$. Then, local perturbations preserving PT and Chiral are 
\begin{align}
    \mathcal{H}_{D}' = m_1 \tilde{\sigma}_0\tilde{\tau}_y + m_2 \tilde{\sigma}_y\tilde{\tau}_z, 
\end{align}
where $m_1$ and $m_2$ represent perturbation strengths or mass terms. The spectrum of $\mathcal{H}_{D}+\mathcal{H}_{D}'$ are given by
\begin{align}
    E = \pm \sqrt{ k_1^2+k_2^2 + m_1^2 + m_2^2 \pm 2\vert m_1k_2 + m_2 k_1\vert},
\end{align}
which indicates that the Dirac nodes are movable but not removable. 
For example, $k_1=510.7 k_x + 76.5 k_y$ and $k_2=-14.7 k_x - 40.9 k_y$ around one Dirac node. Then, turning on the SOC $v_{\text{so}}=15$ meV$\cdot$\AA, we numerically confirm the nodal SC phase with $\Delta_s=3$ meV and $\Delta_o=200$ meV$\cdot$\AA$^2$, shown in Fig.~\ref{fig5} (c), where the logarithm of superconducting gaps are plotted. The eight dark red points are the linear Dirac nodes. Based on the topology-protection argument, the interplay between intra- and inter-orbital pairings for nodal superconductivity is robust against local perturbations. 
Note that our results are different from a previous work~\cite{nakayama_prb_2018}, in which the $d$-wave pairing symmetry induced nodal SC.
In experiments, the nodal gap structure could be detected by measuring the temperature dependence of physical quantities like specific heat and penetration depth at low temperatures. A power law dependence usually indicates the existence of nodal structures (point nodes or line nodes), whereas exponential dependence implies the SC is fully gapped \cite{sigrist_rmp_1991}.

\vspace{10pt}
\noindent
{\bf Applications to superconductors with nematic order.}
In addition to the crystal field splitting, the many-body electron-electron interactions may also lead to orbital hybridization, such as the nematic ordering in the normal states (See Appendix E for details). 
The rotational symmetry reduction could either be from interaction-induced spontaneous symmetry breaking or from explicit symmetry breaking from, say, adding external strain. Then the natural question to ask is whether it is still possible to have an orbital-dependent pairing order characterized by some $\mathbf{d}_o$-vector. Interestingly, we find that the orbital-dependent pairing can coexist with the electronic nematic ordering as long as $\mathbf{d}_o$ is parallel to the $\mathbf{g}_{\mathrm{tot}}$, which is an effective orbital-hybridization vector that also contains the nematic order. This establishes a deep connection between SCs with nematic order and spin-singlet orbital-triplet pairings.
In the following, we study two typical examples. 
\begin{itemize}
\item For case A [two-orbital system], we apply the theory to fit the anisotropic superconducting gap of the hole pocket in the bulk FeSe measured by the quasiparticle interference imaging~\cite{sprau_science_2017}. 
\item For case B [two-valley system], we use a toy model to demonstrate the possible existence of $s+d$-like nematic nodal superconductor in two-valley systems on a honeycomb lattice. We also show the transition between U-shaped and V-shaped quasi-particle density-of-state by tuning the chemical potential.
\end{itemize}

\vspace{10pt}
\noindent
{\bf Case A: Two-orbital model for the bulk FeSe SC.}
We discussed the possible anisotropic $A_{1g}$-type $s$-wave pairing states for the $C_4$-symmetric iron-chalcogenide SCs including fully gapped FeSe thin film and nodal SC in LiFeP and LaFePO. Here we investigate the $C_4$-breaking nematic SC in bulk FeSe. Let us revisit the iron-based SC with a well-established nematic ordering. We consider
$ \mathcal{H}_{\text{int}} = v_1 \hat{n}_1(\mathbf{r})\hat{n}_2(\mathbf{r})$,
where $\hat{n}_i$ is electron density operator for the $i$-atomic orbital. 
If $\langle \hat{n}_1 \rangle \neq \langle \hat{n}_2\rangle$,
$C_n$ ($n>2$) is spontaneously broken down to $C_2$ and we have the nematic order. The intra-orbital interaction does not alter the mean-field results for nematic orders (see details in Appendix E).
The total inter-orbital hybridization contains two parts,
\begin{align}
\mathbf{g}_{\text{tot}}(\bf k) = \mathbf{g}_o(\bf k)+ \mathbf{g}_{\text{nem}},
\end{align}
where $\mathbf{g}_o(\bf k)$ is caused by the crystal field splitting and $\mathbf{g}_{\text{nem}}=(0,0,\Phi)$ is induced due to the nematicity $\Phi=v_1\langle \hat{n}_1 - \hat{n}_2\rangle$, which is momentum-independent if translation symmetry is to be preserved.
Hereafter, we focus on the hole pockets around the $\Gamma$ point to fit the experimental data of superconducting gap functions~\cite{sprau_science_2017}. We will see that even this simplified weak-coupling model, where the coupling between the hole pockets at the $\Gamma$ point and the electron pockets at the $M$ point is ignored, can produce a descent fit the experimental data. A similar result is expected for the electron pockets near the $M$ point. 
Replacing $\mathbf{g}_{o}$ with $\mathbf{g}_{\text{tot}}$ in Eq.~\eqref{Eq::normal_H}, we can still use Eq.~\eqref{eq-tc-lambdao} to investigate the interplay between superconductivity and nematic order, thus the orbital $\mathbf{d}_o$-vector satisfying $\mathbf{d}_o\parallel \mathbf{g}_{\text{tot}}$ leads to the nematic superconductivity. Thus, it generally shows the $A_{1g}$-type $s$-wave spin-singlet orbital-triplet pairings in nematic SCs.

\begin{figure}[t]
	\centering
	\includegraphics[width=\linewidth]{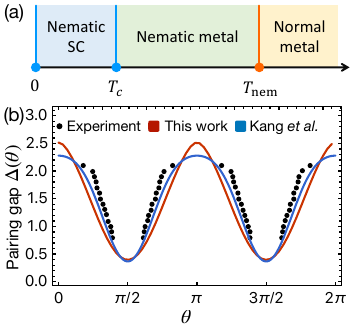}
	\caption{The application to bulk FeSe superconductors with nematicity.
	(a) Schematic phase diagram vs temperature $T$ for normal metal ($T>T_{\text{nem}}$), nematic metal ($T<T_{\text{nem}}$), and nematic superconductivity ($T<T_{c}$). (b) Angular dependence of the superconducting pairing gap: comparison between experiment data (black dots) by Sprau \textit{et al.} Ref.~\cite{sprau_science_2017}, our theory (red line) and the theory proposed by Kang \textit{et al.}~\cite{kang_prl_2018} (blue line). Fitting parameters used for our model: $\Delta_s=2.6$, $\Delta_o=-0.055$ in Eq.~\eqref{eq-pairing-FeSe-fiting}. All the other parameters used are the same~\cite{kang_prl_2018}, including the chemical potential, effective mass, orbital hybridization, and nematic order.
	}
	\label{fig3}
\end{figure}

This scenario can be adopted to study the quasi-two dimensional bulk FeSe, where superconductivity ($T_c\sim8$ K) emerges inside a well-developed nematic phase (transition temperature $T_{\text{nem}}\sim90$ K~\cite{bohmer_jpcm_2017}), shown in Fig.~\ref{fig3} (a). 
For a minimal two-band model \cite{raghu_prb_2008} for the bulk FeSe with $\{d_{xz},d_{yz}\}$-orbitals, $\mathbf{g}_o=(2k_xk_y,0,k_x^2-k_y^2)$ and $\mathbf{g}_{\text{nem}}=(0,0,\Phi)$~\cite{chubukov_prx_2016,kang_prl_2018}. 
Therefore, the nematic orbital $\mathbf{d}_o$-vector with $\mathbf{d}_o \parallel \mathbf{g}_{\text{tot}}$ breaks $C_4$ (see Appendix E for more details). The projected pairing gap function on the large Fermi surface is given by
\begin{align} \label{eq-pairing-FeSe-fiting}
\Delta_{\text{FS}}({\bf k}) = \Delta_s + \Delta_o \sqrt{(-\lambda_o(k_x^2-k_y^2)+\Phi)^2+(2\lambda_ok_xk_y)^2}.
\end{align}
If $\Phi=0$, $\Delta_{\text{FS}}({\bf k})$ is reduced to $\Delta_s + \Delta_o \vert\lambda_o\vert(k_x^2+k_y^2)$ that is in the isotropic limit. The presence of $\Phi$ is the driving force for the anisotropy of $\Delta_{\text{FS}}({\bf k})$. When the nematicity $\Phi$ is strong enough, the orbital $\mathbf{d}_o$-vector will be pinned along the $z$-axis, resulting in the so-called orbital-selective pairing states.
We adopt the realistic parameters for the bulk FeSe SC from Ref.~\cite{kang_prl_2018} to calculate the superconducting gap measured by the quasiparticle interference imaging~\cite{sprau_science_2017}.
In Fig.~\ref{fig3} (b), we show the angular dependence of the pairing gap around the hole pocket at the $\Gamma$-point of FeSe in the presence of nematic order. Our theory provides an equally decent fit to recent experimental data~\cite{sprau_science_2017} as the intra-orbital $s+d$-pairing theory proposed by Kang \textit{et al.}~\cite{kang_prl_2018}, even though our work uses a simplified model without considering the coupling to the other two electron pockets.
Our theory shows more clearly the role of nematic order on the pairing symmetries.
Therefore, the theory developed in this work may alternatively explain the experimental evidence of orbital-selective pairings of the FeSe SC in Refs.~\cite{mcqueen_prl_2009,sprau_science_2017}, and reveal a deep connection between nematic SC and spin-singlet orbital-triplet pairings. It has to be mentioned that here we only focused on the hole pockets around the $\Gamma$ point and discussed the nematicity-induced gap anisotropy around the hole FS. There are other possible mechanisms for gap anisotropy in Fe-based SCs. For example, a previous work~\cite{yu2014orbital} discussed, among other things, the anisotropy/isotropy of the SC gap around the electron pockets at the $M$ point, where the degree of anisotropy depends on the $J_1$-$J_2$ magnetic frustration in the proposed five-orbital $t$-$J_1$-$J_2$ microscopic model.

\begin{figure*}[t]
	\centering
	\includegraphics[width=0.95\textwidth]{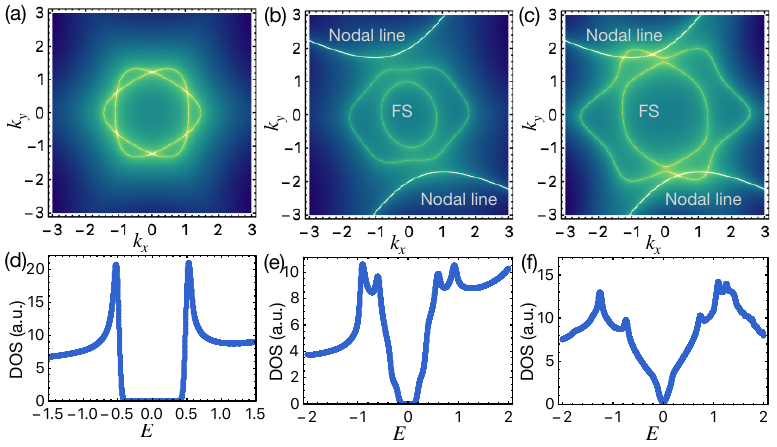}
	\caption{Fermi surfaces (FSs) at the $K_\pm$ valleys and the quasi-particle density of states (DOS). The $C_6$ symmetric FS without inter-valley scattering is shown in (a) and its DOS with an isotropic $s$-wave pairing is given in (d). 
	(b) shows $C_6$-breaking FSs due to the inter-valley scattering, together with the nodal lines of nematic pairing. There are no nodes on the FSs and the corresponding DOS is shown in (e). 
	(c) is similar to (b) but with chemical potential $\mu$ adjusted so that the nodal lines intersect the FSs, hence a V-shaped DOS is obtained as in (f). 
	Parameters used are the following, $\alpha=0.2,t_1=0.15,t_2=0.25,\Delta_o=0.5$. For (a) and (d) $\mu=1.5,g_{\text{int}}=0$; for (b) and (e) $\mu=1.5,g_{\text{int}}=0.7$; for (c) and (f) $\mu=2.7,g_{\text{int}}=0.7$.}
	\label{fig4}
\end{figure*}


\vspace{10pt}
\noindent
{\bf Case B: Two-valley system superconductivity.}
Similar to the two-orbital systems considered above, we discuss in this section superconductivity in two-valley systems, like single layer graphene SC~\cite{wang_prb_2012} or transition metal dichalcogenide (TMD) \cite{hsu2017topological},  where the pairing can be between opposite valleys $K_\pm$.
The spin-singlet pairing is merely characterized by the orbital $\mathbf{d}_o$-vector with $\Delta_s=0$ in Eq.~\eqref{eq-J-2-pairing}.   
For the single-particle Hamiltonian, the inter-valley hopping is naturally forbidden by translational symmetry, namely, $\lambda_o=0$ in Eq.~\eqref{Eq::normal_H}. 
Then, we consider the inter-valley scattering Hamiltonian,
$ \mathcal{H}_{\text{int}} = \sum_{\mathbf{k},\mathbf{k}',\sigma} V(\mathbf{k}-\mathbf{k}') c_{+,\sigma}^\dagger(\mathbf{k}) c_{+,\sigma}(\mathbf{k}') c_{-,\sigma}^\dagger(\mathbf{k}')c_{-,\sigma}(\mathbf{k}) $.
It generates the inter-valley coupling $\mathbf{g}_{\text{int}}$ by defining the order parameter
$\Phi(\mathbf{k}) = \sum_{\mathbf{k}',\sigma} V(\mathbf{k}-\mathbf{k}') \langle c_{+,\sigma}(\mathbf{k}') c_{-,\sigma}^\dagger(\mathbf{k}')  \rangle $ that spontaneously breaks the translational symmetry,
\begin{align}\label{eq-g-int-valley}
\mathbf{g}_{\text{int}}(\mathbf{k})=(g_{\text{int},1}(\mathbf{k}),g_{\text{int},2}(\mathbf{k}),0),
\end{align}
where ${g}_{\text{int},1}(\mathbf{k})  = \text{Re}[\Phi(\mathbf{k}) ]$ and ${g}_{\text{int},2}(\mathbf{k})  = -\text{Im}[\Phi(\mathbf{k}) ]$.
In this case, TRS is $\mathcal{T}=i\tau_1\sigma_2\mathcal{K}$ and IS is $\mathcal{I}=\tau_1\sigma_0$.
The $\mathbf{d}_o$-vector is manifested as $\mathbf{d}_o=(d_1(\mathbf{k}),id_2(\mathbf{k}),0)$ with $d_1(\mathbf{k})=d_1(-\mathbf{k})$ and $d_2(\mathbf{k})=-d_2(-\mathbf{k})$.
Both $d_1(\mathbf{k})$ and $d_2(\mathbf{k})$ are real to preserve TRS.
As for the interaction-induced $\mathbf{g}_{\text{int}}$, $\mathcal{T}$ and $\mathcal{I}$ require $g_{\text{int},1}(\mathbf{k})=g_{\text{int},1}(-\mathbf{k})$ and $g_{\text{int},2}(\mathbf{k})=0$. 
By symmetry, there are two general possibilities.
One is $g_{\text{int},1}(\mathbf{k})=1$, so $C_3\times \mathcal{I}=C_6$ is preserved, and it describes the charge-density-wave order \cite{kekule2016imaging,kekule2021experimental}. 
The other one is $g_{\text{int},1}(\mathbf{k})\in\{k_xk_y,k_x^2-k_y^2\}$ that spontaneously breaks $C_6$ down to $C_2$, forming a nematic order.
This is experimentally possible for the strain-induced Kekul'e distortion (i.e.,~$\sqrt{3}\times\sqrt{3}$ type).

We next discuss superconductivity in the presence of inter-valley couplings, by replacing the $\mathbf{g}_o$-vector  in Eq.~\eqref{Eq::normal_H} with the interaction-induced $\mathbf{g}_{\text{int}}$.
As a result, Eq.~\eqref{eq-tc-lambdao} is still applicable.
It is similar to a recent work~\cite{wolf_arxiv_2021} where the charge order coexists with a sublattice-selective non-unitary pairing state. 

The nematic inter-valley coupling is represented as $g_{\text{int},1}(\mathbf{k})=1+ 2t_1 k_xk_y + t_2 (k_x^2-k_y^2)$, which requires that $\mathbf{d}_o(\mathbf{k})=(1+2t_1 k_xk_y + t_2 (k_x^2-k_y^2),0,0)$ (see Appendix F). Here the normalization factor has been dropped without changing the essential physics. The system is fully gapped if the $s$-wave gap is dominant ($1 \gg \sqrt{t_1^2+t_2^2}$), otherwise, it is a $d$-wave dominant nodal SC ($1 \ll \sqrt{t_1^2+t_2^2}$).

As a concrete toy model, we look at superconductivity on a generic honeycomb lattice with two valleys $K_\pm$, with the Hamiltonian around the two valleys given by,
\begin{align}
\begin{split}
    \mathcal{H}_0(\mathbf{k}) =&\epsilon(\mathbf{k})\tau_0\sigma_0+\alpha(k_x^3-3k_xk_y^2)\tau_3\sigma_0,
\end{split}
\end{align}
where the two-valley basis used here is given by $\psi_{\mathbf{k}}^\dagger=(c_{K_+,\uparrow}^\dagger(\mathbf{k}),  c_{K_+,\downarrow}^\dagger(\mathbf{k}),c_{K_-,\uparrow}^\dagger(\mathbf{k}),c_{K_-,\downarrow}^\dagger(\mathbf{k}))$ and $\epsilon(\mathbf{k})$ takes the same form as in Eq.~(\ref{Eq::normal_H}). The parameter $\alpha$ determines the $C_3$ anisotropy of the continuum model around each valley. This Hamiltonian was used as an effective model~\cite{you2019superconductivity} to study twisted bilayer graphene. 

Including the inter-valley scattering effects, the one-band model is given by
\begin{align}
\begin{split}
    \mathcal{H}(\mathbf{k}) =&\epsilon(\mathbf{k})\tau_0\sigma_0+\alpha(k_x^3-3k_xk_y^2)\tau_3\sigma_0 \\
    &+ \lambda_{\text{int}}[\mathbf{g}_{\text{int}}({\bf k})\cdot\bm{\tau}]\sigma_0,
\end{split}
\end{align}
where the $\lambda_{\text{int}}$ determines the strength of the inter-valley scattering.  
In the absence of inter-valley scattering ($\lambda_{\text{int}}=0$), the Fermi surfaces (FSs) around the two $K_{\pm}$ valleys are plotted in Fig.~\ref{fig4} (a). 
As expected, with a fully symmetric $s$-wave pairing characterized by $\mathbf{d}_o=(1,0,0)$, a fully gapped or U-shaped quasi-particle density-of-states (DOS) is obtained and shown in Fig.~\ref{fig4} (d). 
Then we include the aforementioned inter-valley scattering $\mathbf{g}_{\text{int}}$ that breaks $C_6$ down to $C_2$.
As a result, our theory implies that the effective nematicity generated will favor a nematic pairing characterized by $\mathbf{d}_o\parallel \mathbf{g}_{\text{int}}$. Consider the generic form $\mathbf{d}_o=\mathbf{g}_{\text{int}}=(1+2t_1k_xk_y+t_2(k_x^2-k_y^2),0,0)$, the resulting $C_6$-breaking FS are shown in Fig.~\ref{fig4} (b) and (c), where the nodal lines of the pairing are also shown. By tuning the chemical potential $\mu$, the FSs and nodal lines can go from non-intersecting in Fig.~\ref{fig4} (b) to intersecting in Fig.~\ref{fig4} (c), leading to the corresponding evolution from the gapped U-shaped DOS in Fig.~\ref{fig4} (e) to the gapless V-shaped DOS in Fig.~\ref{fig4} (f).

\vspace{10pt}
\noindent
{\bf Discussions}\\
We briefly discuss the difference between our theory and the previous studies~\cite{fernandes_arcmp_2019} for nematic SCs. One example is a pairing state belonging to a 2D irreducible representation (Irrep), e.g., the $E$-pairing in Cu or Nb-doped Bi$_2$Se$_3$~\cite{fu_prb_2014,matano_np_2016} and UPt$_3$~\cite{sauls_aip_1994,strand_science_2010}. A real order parameter vector $(\Delta_{E,1},\Delta_{E,2})$ spontaneously breaks C$_3$, leading to nematic superconductivity.
Alternatively, a nematic SC can be formed by mixing two different 1D-Irrep-pairing channels.
In FeSe~\cite{Fernandes_prl_2013,kang_prl_2014}, the nematic order breaks the C$_4$ down to C$_2$, which mixes the $s$-wave and $d$-wave pairing channels.
However, $T_c$ of the $(s+d)$ orbital-independent pairing state could be generally affected by increasing nematicity, because of the significant change in the density of states at the Fermi energy. In our theory, the $(s+d)$-like nematic $\mathbf{d}_o$-vector coexists with the nematic order, so $T_c$ is almost unaffected by increasing nematicity. Therefore, it may help to distinguish our results from previous proposals in experiments, where one may use the chemical or physical pressures to tune the nematicity and measure $T_c$ as a function of pressure~\cite{matsuura_nc_2017}. Nevertheless, more efforts are necessary to test the results established in this work for nematic SCs.

\begin{figure}[t]
	\centering
	\includegraphics[width=\linewidth]{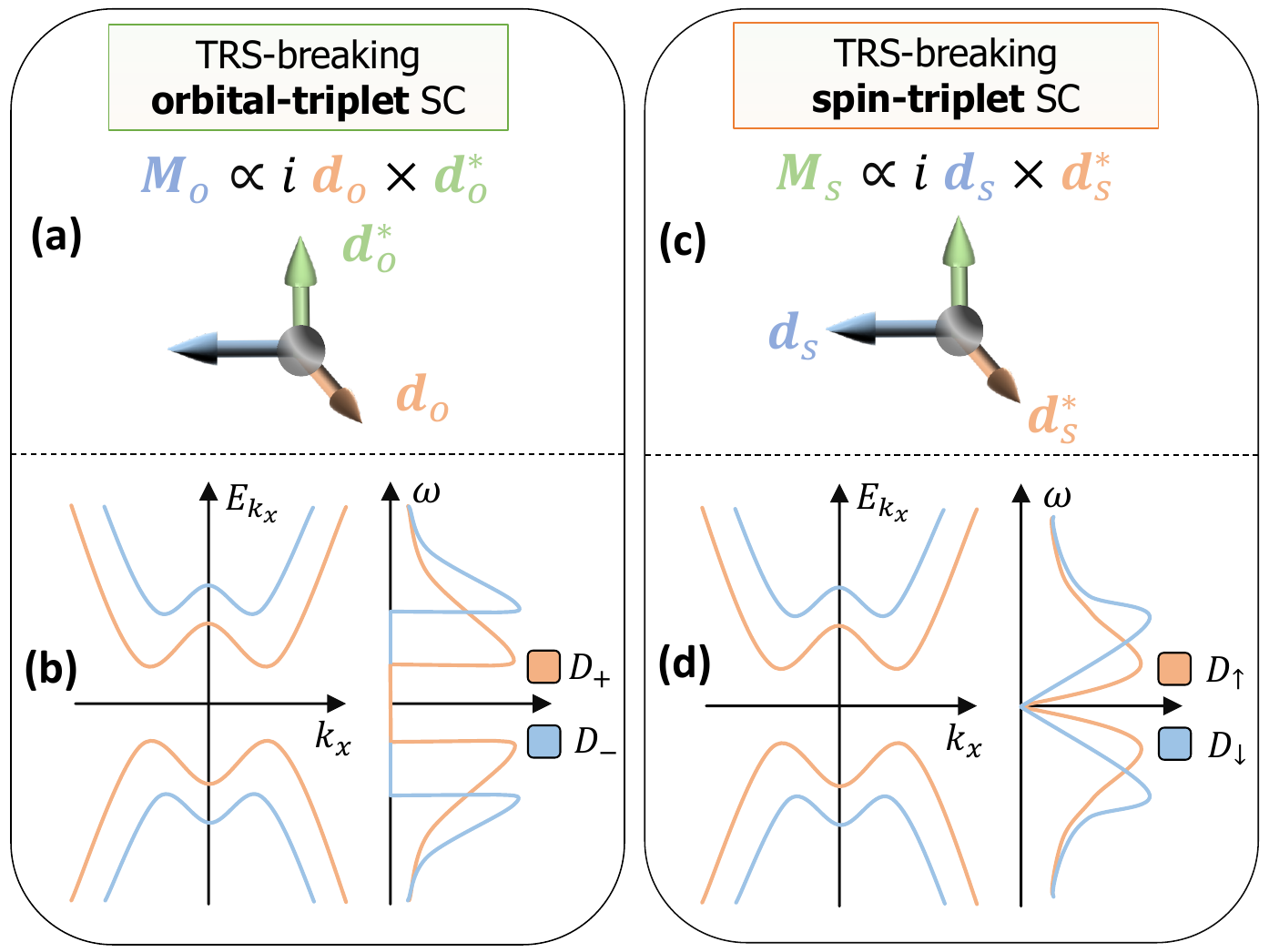}
	\caption{Schematic diagrams for the TRS-breaking effects for non-unitary orbital-triplet superconductors (SCs) and spin-triplet SCs. 
	As for orbital-triplet SCs, (a) shows a complex orbital $\mathbf{d}_o$-vector that spontaneously breaks TRS and results in the TRS-breaking orbital-polarization with $\mathbf{M}_o \propto i\mathbf{d}_o\times\mathbf{d}_o^\ast$; and (b) shows the quasi-particle spectrum along $k_x$ and the orbital-polarized density of states (DOS) $D_{\pm}$ with $\vert\pm\rangle$ representing $\vert 1\rangle\pm i \vert 2\rangle$. 
	For spin-triplet SCs, (c) shows the superconductivity-induced spontaneous spin-polarization with $\mathbf{M}_s \propto i\mathbf{d}_s\times\mathbf{d}_s^\ast$; and (d) shows the two distinct gaps of the quasi-particle spectrum along $k_x$ and the spin-polarized density of state $\mathcal{D}_{\sigma}$ with $\sigma=\{\uparrow,\downarrow\}$. The gapped spectrum is plotted for $k_z\neq 0$ and the node in DOS profile is due to the nodal line at $k_z=0$. 
	}
	\label{fig1}
\end{figure}

In addition to the above discussions for the time-reversal-invariant superconducting states, we also comment on the effects of the spontaneous TRS-breaking, where a complex orbital $\mathbf{d}_o$-vector generates the orbital orderings as $\mathbf{M}_o=-i\gamma_1/\alpha_M(\mathbf{d}\times \mathbf{d}^\ast)$, of which only the $y$-component breaks TRS, as illustrated in Fig.~\ref{fig1} (a).
Alternatively, the corresponding quasi-particle spectrum in Fig.~\ref{fig1} (b) shows the two distinct gaps, similar to the range given by Eq.~\eqref{eq-sc-gap-complex-d}.
More explicitly, we schematically plot the atomic orbital-polarized density of states (DOS) by defining $\vert \pm\rangle= \vert 1\rangle + i \vert 2\rangle$ for complex orbitals, where $D_+\neq D_-$ at finite energy clearly indicates that the DOS is orbital-polarized, which is consistent with the Ginzburg-Landau theory, shown in Appendix B. Moreover, we also find that the orbital-spin conversion would lead to the spin-polarized DOS~\cite{hu_arxiv_xxx}.

The above result for orbital-triplet pairings is similar to the superconducting gaps for non-unitary spin-triplet SCs~\cite{sigrist_rmp_1991}. By symmetry, the Gizburg-Landau free energy is the same. To show the similarity, for the single-band spin-triplet SCs~\cite{leggett_rmp_1975}, the spin-triplet pairing potential is generally given by $\hat{\Delta}(\mathbf{k})=\Delta_0 [\mathbf{d}_s(\mathbf{k})\cdot \bm{\sigma}](i\sigma_2)$, where $\Delta_0$ is the pairing strength and $\bm{\sigma}$ are Pauli matrices in the spin subspace. 
Due to the Fermi statistics, the spin $\mathbf{d}_s(\mathbf{k})$-vector has to satisfy $\mathbf{d}_s(\mathbf{k})=-\mathbf{d}_s(-\mathbf{k})$.
The $\mathbf{d}_s$-vector formalism is firstly developed for He$^3$ superfluid~\cite{Salomaa_RMP1987}.
And it also occurs in noncentrosymmetric SCs, the spin $\mathbf{d}_s(\mathbf{k})$-vector is usually pinned along a certain crystal axis since 
superconductivity is non-suppressed only for $\mathbf{d}_s(\mathbf{k})\parallel \mathbf{g}_s(\mathbf{k})$, where $\mathbf{g}_s(\mathbf{k})$ represents the Rashba spin-orbit coupling (SOC)~\cite{frigeri_prl_2004,ramires_prb_2018}.
Besides, there is intrinsic spontaneous spin-polarization induced by the non-unitary pairing, $\mathbf{d}_s(\mathbf{k})=k_z(1,-i\eta_0,0)$ with real $\eta_0$. 
Fig.~\ref{fig1} (c) shows the spin expectation value of the Cooper pairs ($\mathbf{M}_s \propto i\mathbf{d}_s^\ast(\mathbf{k})\times\mathbf{d}_s(\mathbf{k})=2\eta_0k_z^2\vec{e}_z$).
It is an equal-spin pairing so that $\sigma_3$ is conserved, and non-zero $\mathbf{M}_s$ leads to two distinct superconducting gaps of the quasi-particle spectrum \cite{sigrist_AIP_2005}, shown in Fig.~\ref{fig1} (d).
In addition, the density of states (DOS) is spin-polarized, namely, $D_\uparrow\neq D_\downarrow$ at finite energy $\omega$, as illustrated in Fig.~\ref{fig1} (d).

To summarize, we have derived a general weak-coupling criterion to investigate the spin-singlet orbital-triplet pairings in nematic SCs. 
For technical convenience, we adopt the orbital $\mathbf{d}_o$-vector to describe the spin-singlet orbital-dependent pairing states and the $\mathbf{g}_o$-vector for the orbital hybridizations. The main results of this work include, first, we demonstrate that an orbital $\mathbf{d}_o$-vector that is parallel with $\mathbf{g}_o$-vector for orbital hybridizations is possible to be realized in real superconducting materials. Second, the interplay between intra-orbital and orbital-dependent pairings that belong to the same symmetry representation can explain the observation of robust Dirac nodes in the quasi-2D iron-based SCs. Remarkably, we find that $\mathbf{d}_o$-vectors could even coexist with many-body interaction-induced nematic orders or charge-density-wave orders when $\mathbf{d}_o\propto \mathbf{g}_{\text{tot}}=\mathbf{g}_o+\mathbf{g}_{\text{nem}}$ (or $\mathbf{g}_{\text{int}}$). Moreover, our theory discovers the important role of nematic orders in SC pairing symmetry, which builds a possible bridge between repulsive interaction-induced nematic orders and nematic superconductivity and also reveals a deep connection between spin-singlet orbital-triplet pairings in nematic SCs. Our results may be helpful in understanding the nematic superconductivity in bulk FeSe.
Our work will motivate more theoretical and experimental efforts to search for spin-singlet orbital-triplet SCs, even for topological superconductivity, which might contribute to further understanding the effects of spontaneous symmetry breaking on superconductivity.

\vspace{10pt}
\noindent
{\bf Methods}\\
In this section, we present the derivation for the main result Eq.~(\ref{eq-tc-lambdao}), which is first order in $\lambda_o$, by solving the linearized gap equation. The second order results are presented in Appendix C. 
The general $\mathbf{k}\cdot\mathbf{p}$ normal Hamiltonian considered in the main text reads,
\begin{equation}
\begin{split}
\mathcal{H}_0(\mathbf{k}) = \epsilon(\mathbf{k})\tau_0\sigma_0 + \lambda_o(\mathbf{g}_o(\mathbf{k})\cdot\bm{\tau})\sigma_0,
\end{split}
\end{equation}
where the electronic basis is made of $\{1,2\}$-orbitals $\Psi_{\mathbf{k}}^\dagger=(c_{1,\uparrow}^\dagger(\mathbf{k}),  c_{1,\downarrow}^\dagger(\mathbf{k}),c_{2,\uparrow}^\dagger(\mathbf{k}),c_{2,\downarrow}^\dagger(\mathbf{k}))$,
$\epsilon(\mathbf{k}) = (k_x^2+k_y^2)/2m -\mu$ is the band energy measured relative to the chemical potential $\mu$, 
$\lambda_o$ represents the orbital hybridization and $\mathbf{g}_o(\mathbf{k})=(g_1(\mathbf{k}),g_2(\mathbf{k}),g_3(\mathbf{k}))$.
The TRS $\mathcal{T}=i\sigma_2\tau_0\mathcal{K}$ requires $g_{1,3}(\mathbf{k})=g_{1,3}(-\mathbf{k})$ and $g_{2}(\mathbf{k})=-g_{2}(-\mathbf{k})$.
It leads that
\begin{align}
\mathbf{g}_o(\mathbf{k})\cdot\bm{\tau} = \left\lbrack  \mathbf{g}_o(-\mathbf{k})\cdot\bm{\tau}  \right\rbrack^\ast.
\end{align}
Besides, we set $\lambda_o>0$ without loss of generality.
The Matsubara Green's function for electrons is $G_e(\mathbf{k},i\omega_n)=[i\omega_n-\mathcal{H}_0(\mathbf{k})]^{-1}$ and that for holes is $G_h(\mathbf{k},i\omega_n)=-G_e^\ast(\mathbf{k},i\omega_n)$. Here $\beta=1/k_BT$ and $\omega_n=(2n+1)\pi/\beta$ with $n$ integer. 
Therefore,
\begin{widetext}
\begin{align}
G_e(\mathbf{k},i\omega_n) &= \frac{\mathcal{P}_-(\mathbf{k})}{i\omega_n  -\epsilon(\mathbf{k}) +\lambda_o\vert\mathbf{g}_o(\mathbf{k})\vert } + \frac{\mathcal{P}_+(\mathbf{k})}{i\omega_n -\epsilon(\mathbf{k}) -\lambda_o\vert\mathbf{g}_o(\mathbf{k})\vert } \triangleq G_e^-(\mathbf{k},i\omega_n) \mathcal{P}_-(\mathbf{k})  + G_e^+(\mathbf{k},i\omega_n) \mathcal{P}_+(\mathbf{k}) ,\\
G_h(-\mathbf{k},i\omega_n) &= \frac{\mathcal{P}_-(\mathbf{k})}{i\omega_n  +\epsilon(\mathbf{k}) -\lambda_o\vert\mathbf{g}_o(\mathbf{k})\vert } + \frac{\mathcal{P}_+(\mathbf{k})}{i\omega_n +\epsilon(\mathbf{k}) +\lambda_o\vert\mathbf{g}_o(\mathbf{k})\vert } \triangleq G_h^-(\mathbf{k},i\omega_n) \mathcal{P}_-(\mathbf{k})  + G_h^+(\mathbf{k},i\omega_n) \mathcal{P}_+(\mathbf{k}) ,
\end{align}
\end{widetext}
where $\mathcal{P}_\pm(\mathbf{k})=\frac{1}{2}( 1\pm \hat{\mathbf{g}}_o(\mathbf{k})\cdot\bm{\tau} )$ with $\hat{\mathbf{g}}_o(\mathbf{k})={\mathbf{g}}_o(\mathbf{k})/\vert {\mathbf{g}}_o(\mathbf{k})\vert$.
Here $G_e^\pm(\mathbf{k},i\omega_n) = \tfrac{1}{i\omega_n  -\epsilon(\mathbf{k}) \mp \lambda_o\vert\mathbf{g}_o(\mathbf{k})\vert }$ and $G_h^\pm(\mathbf{k},i\omega_n) = \tfrac{1}{i\omega_n  +\epsilon(\mathbf{k}) \pm \lambda_o\vert\mathbf{g}_o(\mathbf{k})\vert }$. 
We expand the attractive interactions as 
\begin{equation}
\begin{split}
V^{s_1a,s_2b}_{s_1'a',s_2'b'}(\mathbf{k},\mathbf{k}') =& -\sum_{\Gamma,l}v_0^\Gamma \lbrack \mathbf{d}_o^{\Gamma,l}(\mathbf{k})\cdot\bm{\tau}i\sigma_2 \rbrack_{s_1a,s_2b} \\
&\times \lbrack \mathbf{d}_o^{\Gamma,l}(\mathbf{k}')\cdot\bm{\tau}i\sigma_2 \rbrack_{s_1'a',s_2'b'},
\end{split}
\end{equation}
where $v_0^\Gamma>0$ is the interaction strength of the irreducible representation channel $\Gamma$ of the crystalline group, and $l=1,2,...,\mathrm{Dim}\ \Gamma$. Each pairing channel $\Gamma$ gives rise to a SC critical temperature $T_c^\Gamma$, and the actual transition temperature of the system is given by the largest of these critical temperatures. In our work, we main focus on the case where $\mathrm{Dim}\ \Gamma=1$, which is sufficient for the applications discussed in the main text. 
The coupling between orbital-dependent pairings and orbital-independent pairings will be discussed in details later.
The transition temperature $T_c^\Gamma$ of orbital-dependent pairing channels is calculated by solving the linearized gap equation,
\begin{equation}
\begin{split}
\Delta_{s_1,s_2}^{a,b}(\mathbf{k}) =& -\frac{1}{\beta} \sum_{\omega_n}\sum_{s_1'a',s_2'b'} V^{s_1a,s_2b}_{s_1'a',s_2'b'}(\mathbf{k},\mathbf{k}')\\
&\times 
\left\lbrack G_e(\mathbf{k}',i\omega_n) \hat{\Delta}(\mathbf{k}') G_h(-\mathbf{k}',i\omega_n) \right\rbrack_{s_1'a',s_2'b'},
\end{split}
\end{equation}
which is reduced to $v_0^\Gamma\chi^\Gamma(T)-1 = 0$ with the superconductivity susceptibility $\chi^\Gamma(T)$ in the channel $\Gamma$ defined as,
\begin{widetext}
\begin{align}
\chi^\Gamma(T) &= -\frac{1}{\beta}\sum_{\mathbf{k},\omega_n} \text{Tr}\left\lbrack (\mathbf{d}_o^{\Gamma}(\mathbf{k})\cdot\bm{\tau}i\sigma_2)^\dagger G_e(\mathbf{k},i\omega_n) (\mathbf{d}_o^{\Gamma}(\mathbf{k})\cdot\bm{\tau}i\sigma_2) G_h(-\mathbf{k},i\omega_n) \right\rbrack, \\
&=-\frac{2}{\beta}\sum_{\mathbf{k},\omega_n} \sum_{\alpha,\beta} G_e^\alpha(\mathbf{k},i\omega_n) G_h^\beta(\mathbf{k},i\omega_n)\times \text{Tr}\left\lbrack (\mathbf{d}_o^{\Gamma}(\mathbf{k})\cdot\bm{\tau})^\dagger \mathcal{P}_\alpha(\mathbf{k}) (\mathbf{d}_o^{\Gamma}(\mathbf{k})\cdot\bm{\tau}) \mathcal{P}_\beta(\mathbf{k})  \right\rbrack,
\end{align}
\end{widetext}
where $\alpha,\beta\in\{+,-\}$.
For notional simplicity, the superscript $\Gamma$ will be dropped when there is no danger of confusion. Firstly, let us calculate the trace part.  In the following calculate, we will use
\begin{widetext}
\begin{align}
\begin{split}
&\text{Tr}\left\lbrack (\mathbf{d}_o(\mathbf{k})\cdot\bm{\tau})^\dagger \mathcal{P}_+(\mathbf{k}) (\mathbf{d}_o(\mathbf{k})\cdot\bm{\tau}) \mathcal{P}_+(\mathbf{k})  \right\rbrack \\
&\quad+ \text{Tr}\left\lbrack (\mathbf{d}_o(\mathbf{k})\cdot\bm{\tau})^\dagger \mathcal{P}_-(\mathbf{k}) (\mathbf{d}_o(\mathbf{k})\cdot\bm{\tau}) \mathcal{P}_-(\mathbf{k})  \right\rbrack \\
&=\left( \mathbf{d}_o^\ast(\mathbf{k})\cdot\mathbf{d}_o(\mathbf{k}) \right)  
+ 2\left( \mathbf{d}_o^\ast(\mathbf{k})\cdot\hat{\mathbf{g}}_o(\mathbf{k}) \right)\left( \mathbf{d}_o(\mathbf{k})\cdot\hat{\mathbf{g}}_o(\mathbf{k}) \right) \left( \mathbf{d}_o^\ast(\mathbf{k})\cdot\mathbf{d}_o(\mathbf{k}) \right) \left( \hat{\mathbf{g}}_o(\mathbf{k})\cdot\hat{\mathbf{g}}_o(\mathbf{k}) \right). 
\end{split}
\end{align}
And,
\begin{align}
\begin{split}
&\text{Tr}\left\lbrack (\mathbf{d}_o(\mathbf{k})\cdot\bm{\tau})^\dagger \mathcal{P}_+(\mathbf{k}) (\mathbf{d}_o(\mathbf{k})\cdot\bm{\tau}) \mathcal{P}_-(\mathbf{k})  \right\rbrack  + \text{Tr}\left\lbrack (\mathbf{d}_o(\mathbf{k})\cdot\bm{\tau})^\dagger \mathcal{P}_-(\mathbf{k}) (\mathbf{d}_o(\mathbf{k})\cdot\bm{\tau}) \mathcal{P}_+(\mathbf{k})  \right\rbrack \\
=&\left( \mathbf{d}_o^\ast(\mathbf{k})\cdot\mathbf{d}_o(\mathbf{k}) \right)  
- 2\left( \mathbf{d}_o^\ast(\mathbf{k})\cdot\hat{\mathbf{g}}_o(\mathbf{k}) \right)\left( \mathbf{d}_o(\mathbf{k})\cdot\hat{\mathbf{g}}_o(\mathbf{k}) \right) 
+ \left( \mathbf{d}_o^\ast(\mathbf{k})\cdot\mathbf{d}_o(\mathbf{k}) \right) \left( \hat{\mathbf{g}}_o(\mathbf{k})\cdot\hat{\mathbf{g}}_o(\mathbf{k}) \right). 
\end{split}
\end{align}

Therefore, we arrive at
\begin{align}
\begin{split}
&\text{Tr}\left\lbrack (\mathbf{d}_o(\mathbf{k})\cdot\bm{\tau})^\dagger \mathcal{P}_\alpha(\mathbf{k}) (\mathbf{d}_o(\mathbf{k})\cdot\bm{\tau}) \mathcal{P}_\beta(\mathbf{k})  \right\rbrack  \\
=&\frac{1}{2}\Big{\lbrack} \left( \mathbf{d}_o^\ast(\mathbf{k})\cdot\mathbf{d}_o(\mathbf{k}) \right) 
+ i\alpha \left( \mathbf{d}_o(\mathbf{k})\cdot \left( \mathbf{d}_o^\ast(\mathbf{k}) \times \hat{\mathbf{g}}_o(\mathbf{k}) \right) \right)  
+ i\beta\left( \mathbf{d}_o^\ast(\mathbf{k})\cdot \left( \mathbf{d}_o(\mathbf{k}) \times \hat{\mathbf{g}}_o(\mathbf{k}) \right) \right) \\
+& \alpha\beta\left(  2\left( \mathbf{d}_o^\ast(\mathbf{k})\cdot\hat{\mathbf{g}}_o(\mathbf{k}) \right)\left( \mathbf{d}_o(\mathbf{k})\cdot\hat{\mathbf{g}}_o(\mathbf{k}) \right)  - 
\left( \mathbf{d}_o^\ast(\mathbf{k})\cdot\mathbf{d}_o(\mathbf{k}) \right) \left( \hat{\mathbf{g}}_o(\mathbf{k})\cdot\hat{\mathbf{g}}_o(\mathbf{k}) \right) 
\right) \Big{\rbrack}.
\end{split}
\end{align}
Then we have 
\begin{equation}
\label{eq:chi_T}
\begin{split}
  \chi(T) &=-\frac{1}{\beta}\sum_{\mathbf{k},\omega_n} \sum_{\alpha,\beta} G_e^\alpha(\mathbf{k},i\omega_n) G_h^\beta(\mathbf{k},i\omega_n)
  \Big{\lbrack} \left( \mathbf{d}_o^\ast(\mathbf{k})\cdot\mathbf{d}_o(\mathbf{k}) \right) 
+ i\alpha \left( \mathbf{d}_o(\mathbf{k})\cdot \left( \mathbf{d}_o^\ast(\mathbf{k}) \times \hat{\mathbf{g}}_o(\mathbf{k}) \right) \right)  
+ i\beta\left( \mathbf{d}_o^\ast(\mathbf{k})\cdot \left( \mathbf{d}_o(\mathbf{k}) \times \hat{\mathbf{g}}_o(\mathbf{k}) \right) \right) \\
+& \alpha\beta\left(  2\left( \mathbf{d}_o^\ast(\mathbf{k})\cdot\hat{\mathbf{g}}_o(\mathbf{k}) \right)\left( \mathbf{d}_o(\mathbf{k})\cdot\hat{\mathbf{g}}_o(\mathbf{k}) \right)  - 
\left( \mathbf{d}_o^\ast(\mathbf{k})\cdot\mathbf{d}_o(\mathbf{k}) \right) \left( \hat{\mathbf{g}}_o(\mathbf{k})\cdot\hat{\mathbf{g}}_o(\mathbf{k}) \right) 
\right) \Big{\rbrack}.
\end{split}
\end{equation}
\end{widetext}
Next, we calculate the integration for $\sum_{\mathbf{k},\omega_n}$ by using,
\begin{align}
\sum_{\mathbf{k},\omega_n} \to \frac{N_0}{4}\int_{-\omega_D}^{+\omega_D} d\epsilon \int_{S}\frac{d\Omega}{2\pi} \, \sum_{\omega_n},
\end{align}
where $N_0$ is the density of states at Fermi surface and $\Omega$ is the solid angle of $\mathbf{k}$ on Fermi surfaces. Then,
\begin{equation}
\begin{split}
\label{sm-eq-first-order-GeGh}
&-\frac{N_0}{\beta} \int_{-\omega_D}^{+\omega_D}d\epsilon\int_{S}\frac{d\Omega}{2\pi}\sum_{\omega_n}  G_e^+(\mathbf{k},i\omega_n) G_h^+(\mathbf{k},i\omega_n)\\
&\quad = -\frac{N_0}{\beta} \int_{-\omega_D}^{+\omega_D}d\epsilon\int_{S}\frac{d\Omega}{2\pi} \sum_{\omega_n} G_e^-(\mathbf{k},i\omega_n) G_h^-(\mathbf{k},i\omega_n)\\
&\quad \equiv \chi_0(T),
\end{split}
\end{equation}

On one hand, 
\begin{equation}
\begin{split}
\int_{-\omega_D}^{+\omega_D}&d\epsilon\sum_{\omega_n}  G_e^+(\mathbf{k},i\omega_n) G_h^+(\mathbf{k},i\omega_n)\\
&= \int_{-\omega_D}^{+\omega_D}d\epsilon\sum_{\omega_n}\frac{1}{i\omega_n+\epsilon}\frac{1}{i\omega_n-\epsilon}\\
&=\beta\int_{-\omega_D}^{+\omega_D}d\epsilon\frac{\tanh{\frac{\beta\epsilon}{2}}}{2\epsilon}\\
&=\beta\int_0^{\beta\omega_D/2}dx\frac{\tanh{x}}{x}\approx \beta\ln\left( \tfrac{2e^\gamma \omega_D}{\pi k_B T}\right),
\end{split}
\end{equation}
where the approximation is done at low temperature when $\beta\to\infty$.

On the other hand, we could find a series representation for $\chi_0$, which also applies to the case where $\lambda_o\neq 0$, so that $\chi_0\equiv \chi(\lambda_o=0)$ and $\chi(\lambda_o\neq 0)$ can be related by a simple relation. The way to do it is to perform the integration in $\epsilon$ first. More precisely,
\begin{equation}
\begin{split}
    &\int_{-\omega_D}^{+\omega_D}
    d\epsilon\sum_{\omega_n}G_e^+(\mathbf{k},i\omega_n) G_h^+(\mathbf{k},i\omega_n) \\
    &= \int_{-\omega_D}^{+\omega_D}\sum_{\omega_n}\frac{1}{i\omega_n+\epsilon}\frac{1}{i\omega_n-\epsilon}\\
    &=2\mathrm{Re}\sum_{n\geq 0}\int_{-\omega_D}^{\omega_D}d\epsilon\frac{1}{i\omega_n+\epsilon}\frac{1}{i\omega_n-\epsilon}\\
    &=2\beta\mathrm{Re}\sum_{n\geq 0}\int_{-\beta\omega_D}^{\beta\omega_D}d\epsilon\frac{1}{i2\pi(n+1/2)+\epsilon}\frac{1}{i2\pi(n+1/2)-\epsilon}\\
   & \approx 2\beta\mathrm{Re}\sum_{n\geq 0}\int_{-\infty}^{\infty}d\epsilon\frac{1}{i2\pi(n+1/2)+\epsilon}\frac{1}{i2\pi(n+1/2)-\epsilon}\\
    &=\beta \mathrm{Re}\sum_{n\geq 0}\frac{1}{n+1/2},
\end{split}
\end{equation}
where the low temperature limit is again assumed and the integration is done using residue theorem. In the same spirit, we have,
\begin{equation}
\begin{split}
    \int_{-\omega_D}^{+\omega_D}&d\epsilon\sum_{\omega_n}  G_e^+(\mathbf{k},i\omega_n) G_h^-(\mathbf{k},i\omega_n)\\
    &=\beta \mathrm{Re}\sum_{n\geq 0}\frac{1}{n+1/2+i\frac{\lambda_o|\bf g_o(\bf k)|}{2\pi k_BT}},
\end{split}
\end{equation}

Now by introducing the digamma function defined on the complex plane,
\begin{equation}
    \psi^{(0)}(z)=-\gamma+\sum_{n\geq 0}(\frac{1}{n+1}-\frac{1}{n+z}),
\end{equation}
we have the following relation,
\begin{equation}
\begin{split}
    \int_{-\omega_D}^{+\omega_D}&d\epsilon\sum_{\omega_n}  G_e^+(\mathbf{k},i\omega_n) G_h^-(\mathbf{k},i\omega_n)\\
    &-\int_{-\omega_D}^{+\omega_D}d\epsilon\sum_{\omega_n}  G_e^+(\mathbf{k},i\omega_n) G_h^+(\mathbf{k},i\omega_n) \\
    &=\beta\text{Re}\lbrack \psi^{(0)}(\tfrac{1}{2}) - \psi^{(0)}(\tfrac{1}{2}+i\tfrac{\lambda_o\vert\mathbf{g}(\mathbf{k})\vert}{2\pi k_BT}) \rbrack\\
    &\equiv \beta\mathcal{C}_0(T),
\end{split}
\end{equation}
where $\chi_0(T)=N_0\ln\left( \tfrac{2e^\gamma \omega_D}{\pi k_B T} \right)$, $\gamma=0.57721\cdots$ is the Euler-Mascheroni constant and $\omega_D$ is the Debye frequency.

Therefore,
\begin{equation}
\begin{split}
-\frac{N_0}{\beta} &\int_{-\omega_D}^{+\omega_D}d\epsilon\int_{S}\frac{d\Omega}{2\pi} \sum_{\omega_n} G_e^-(\mathbf{k},i\omega_n) G_h^+(\mathbf{k},i\omega_n)\\ 
&= -\frac{N_0}{\beta} \int_{-\omega_D}^{+\omega_D}d\epsilon\int_{S}\frac{d\Omega}{2\pi} \sum_{\omega_n} G_e^+(\mathbf{k},i\omega_n) G_h^-(\mathbf{k},i\omega_n)\\
&= \chi_0(T) + N_0\int_{S}\frac{d\Omega}{2\pi}\mathcal{C}_0(T).
\end{split}
\end{equation}
Now we can proceed to calculate $\chi(T)$ given in Eq.~(\ref{eq:chi_T}),

\begin{align}
\chi(T) &= \chi_0(T)\int_{S}\frac{d\Omega}{2\pi}\left\vert \mathbf{d}_o\cdot\hat{\mathbf{g}}_o \right\vert^2 \\  
&\quad +\chi_0(T)\int_{S}\frac{d\Omega}{2\pi}\left( |\mathbf{d}_o|^2- \left\vert \mathbf{d}_o\cdot\hat{\mathbf{g}}_o \right\vert^2 \right)\\
&\quad +N_0\int_{S}\frac{d\Omega}{2\pi}\mathcal{C}_0(T)\left( |\mathbf{d}_o|^2- \left\vert \mathbf{d}_o\cdot\hat{\mathbf{g}}_o \right\vert^2 \right)
 \\
&=\chi_0(T)+N_0\int_{S} \frac{d\Omega}{2\pi} \, \mathcal{C}_0(T) \left( \left\vert \mathbf{d}_o \right\vert^2 -  \left\vert \mathbf{d}_o\cdot\hat{\mathbf{g}}_o \right\vert^2 \right).
\label{sm-eq-chi0-do}
\end{align}
In the calculation, we use normalized gap functions with  $\int_{S}\frac{d\Omega}{2\pi} \, \mathbf{d}_o^\ast\cdot\mathbf{d}_o=1 $. It leads to,
\begin{align}\label{sm-tc-lambda0}
\ln\left(\frac{T_c}{T_{c0}} \right) =\int_{S} \frac{d\Omega}{2\pi} \, \mathcal{C}_0(T_{c}) \left( \left\vert \mathbf{d}_o \right\vert^2 -  \left\vert \mathbf{d}_o\cdot\hat{\mathbf{g}}_o \right\vert^2 \right),
\end{align}
where $T_{c0}$ is $T_c$ for $\lambda_o=0$ case by solving $v_0\chi_0(T_{c0})=1$. This is the Eq.~(\ref{eq-tc-lambdao}) in the main text.
In general, the right-hand side of Eq.~\eqref{sm-tc-lambda0} suppress $T_c$.
It clearly indicates that $T_c$ would not be suppressed by orbital hybridization once $\mathbf{d}_o\parallel \mathbf{g}_o$ for all $\mathbf{k}$.
So we conclude that the orbital $\mathbf{d}_o$-vector is possible to be stabilized in materials.


\vspace{10pt}
\noindent
{\bf Data availability}\\
The datasets generated during this study are available from the corresponding author upon reasonable request.

\vspace{10pt}
\noindent
{\bf Code availability}\\
The custom codes generated during this study are available from the corresponding author upon reasonable request.

\vspace{10pt}
\noindent
{\bf Acknowledgments}\\
We thank J.~Yu, X.~X. Wu, C.-X.~Liu, D.-C.~Lu, and A.~Kreisel for helpful discussions. We especially acknowledge J.~Yu's careful reading of the manuscript.
L.-H.~Hu acknowledges the support of a DOE grant (DESC0019064) and the Office of Naval Research (Grant No. N00014-18-1-2793).
D.-H.~Xu was supported by the NSFC (under Grant Nos. 12074108 and 12147102).
F.-C.~Zhang is partially supported by NSFC grant No.~11920101005 and No.~11674278, and by the Priority Program of Chinese Academy of Sciences, grant No. XDB28000000. F.-C.~Zhang was partially supported by Chinese Academy of Sciences under contract No. JZHKYPT-2021-08.

\vspace{10pt}
\noindent
{\bf Author contributions}\\
L.-H. H and F.-C. Z. supervised the project.
M. Z. performed numerical calculations
with the help of L.-H. H. 
All authors contributed to analyzing the
data and writing the manuscript.

\vspace{10pt}
\noindent
{\bf Competing interests}\\
The authors declare no competing interests.

\vspace{10pt}
\def\bibsection{\noindent{\bf\refname}}
\bibliography{ref}

\clearpage
\appendix

\begin{widetext}
\section{Two definitions for the orbital $\mathbf{d}_o$-vector}
In the main text, we take the general pairing potential of a two-orbital SC,
\begin{align}\label{sm-eq-singlet-pairing-tot}
\hat{\Delta}_{tot}(\mathbf{k})= (\Delta_s\Psi_s(\mathbf{k})\tau_0+\Delta_o(\mathbf{d}_o(\mathbf{k})\cdot\bm{\tau}))(i\sigma_2),
\end{align} 
where $\Delta_s$ and $\Delta_o$ are pairing strengths in orbital-independent and orbital-dependent channels, respectively. 
Here $\bm{\tau}$ are Pauli matrices acting on the orbital subspace and $\tau_0$ is a 2-by-2 identity matrix. 
In the absence of band-splitting caused by spin-orbital couplings, the gap function on the Fermi surface is 
\begin{align}
\Delta(\mathbf{k}) = \sqrt{\vert\Delta_s\vert^2 \Psi_s^2(\mathbf{k}) + \vert\Delta_o\vert^2 \vert\mathbf{d}_o(\mathbf{k})\vert^2\pm \vert \mathbf{q}_o\vert  },
\end{align}
where $\mathbf{q}_o= i\vert\Delta_o\vert^2 (\mathbf{d}_o^\ast(\mathbf{k})\times \mathbf{d}_o(\mathbf{k})) + \text{ Re}[\Delta_s^\ast\Delta_o\mathbf{d}_o(\mathbf{k})]$. This expression is mathematically similar to the superconducting gap of non-unitary spin-triplet SCs.

At this point, it is a good place to comment on the other possible way to defining the orbital $\mathbf{d}_o$-vector. Different from the one used in the main text, this definition groups the pairing term into orbital-singlet and orbital-triplet parts. In the form of Eq.~(A1), $\Psi_s(\mathbf{k})$ and $d_o^{1,3}(\mathbf{k})$ are even in $\mathbf{k}$, but $d_o^2(\mathbf{k})$ is odd in $\mathbf{k}$ due to Fermi statistics. By regrouping the terms based on the parity in $\mathbf{k}$, we have 
\begin{align}
\hat{\Delta}(\mathbf{k})=[\Delta_od_o^2(\mathbf{k})\tau_0+(-i\Delta_od_o^3(\mathbf{k}),\Delta_s\Psi_s(\mathbf{k}),i\Delta_od_o^1(\mathbf{k}))\cdot \bm{\tau}](\tau_2i\sigma_2),
\end{align}
which contains $\tilde{\mathbf{d}}_o\cdot \bm{\tau}$ with the new $\tilde{\mathbf{d}}_o$-vector redefined in terms of the original amplitudes and form factors. In this form, the first part is odd in $\mathbf{k}$, which is the orbital-singlet part, and the second part is even in $\mathbf{k}$ and gives orbital-triplet state. Table~\ref{tab:two-d-vector} gives a detailed comparison between the two definitions of the orbital $\mathbf{d}_o$-vector. The spin $\mathbf{d}_s$-vector is also presented for completeness. 
It shows that the definition of orbital ${\mathbf{d}}_o$-vector used in the main text is more convenient to discuss the spontaneous TRS-braking pairing states.

\begin{table}[!htbp]
	\begin{ruledtabular}
		\begin{tabular}{c|c|c|c}
			& orbital $\mathbf{d}_o$  & orbital $\tilde{\mathbf{d}}_o$ & spin $\mathbf{d}_s$\\ \hline
			Pairing potential & $[\Delta_s\Psi_s(\mathbf{k})\tau_0+\Delta_o\mathbf{d}_o(\mathbf{k})\cdot\bm{\tau}]i\sigma_2$ &$ [\tilde{\Delta}_s\tilde{\Psi}_s(\mathbf{k})\tau_0+\tilde{\Delta}_o\tilde{\mathbf{d}}_o(\mathbf{k})\cdot\bm{\tau}]\tau_2i\sigma_2$ & $[\Delta_t\mathbf{d}_s(\mathbf{k})\cdot\bm{\sigma}]i\sigma_2$\\ \hline
			Parity &$\Psi_s(\mathbf{k})$, $d_o^2(\mathbf{k})$ odd; $d_o^{1,3}(\mathbf{k})$ even &$\tilde{\Psi}_s(\mathbf{k})$ odd; $\tilde{\mathbf{d}}_o(\mathbf{k})$ even & $\mathbf{d}_s(\mathbf{k})$ odd\\ \hline
			TRS &\makecell{$\Psi_s^\ast(\mathbf{k})=\Psi_s(\mathbf{k})$;\\ 
				$\mathbf{d}_o^\ast(\mathbf{k})=\mathbf{d}_o(\mathbf{k})$; \\ } & \makecell{$\tilde{\Psi}_s^\ast(\mathbf{k})=\tilde{\Psi}_s(\mathbf{k})$;\\ $\left(\tilde{d}_o^{1,3}(\mathbf{k})\right)^\ast=-\tilde{d}_o^{1,3}(\mathbf{k})$;\\ $\left(\tilde{d}_o^2(\mathbf{k})\right)^\ast=\tilde{d}_o^2(\mathbf{k})$}& $\mathbf{d}_s^*(\mathbf{k})=\mathbf{d}_s(\mathbf{k})$ \\ \hline
			TRS breaking & complex $\mathbf{d}_o$ & $\tilde{d}_o^{1,3}(\mathbf{k})$ real or $\tilde{d}_o^2(\mathbf{k})$ complex & complex $\mathbf{d}_s$ \\ \hline
			spontaneous AOP/SP & $\mathbf{M}_o\propto i\mathbf{d}_o^\ast(\mathbf{k})\times \mathbf{d}_o(\mathbf{k})$ &$\mathbf{M}_o\propto i\tilde{\mathbf{d}}_o^\ast(\mathbf{k})\times \tilde{\mathbf{d}}_o(\mathbf{k})$ & $\mathbf{M}_s\propto i\mathbf{d}_s^\ast(\mathbf{k})\times \mathbf{d}_s(\mathbf{k})$\\ 
			
		\end{tabular}
		\caption{\label{tab:two-d-vector} 
			Comparison between the two possible definitions of the orbital $\mathbf{d}_o$-vector in spin-singlet SCs, together with the spin $\mathbf{d}_s$-vector of spin-triplet SCs. The parity properties are obtained from Fermi statistics. The TRS row gives the transformation properties in order to preserve TRS. Both the atomic orbital polarization (AOP) and the spin polarization (SP) take the same form in terms of their respective $\mathbf{d}$-vectors.}
	\end{ruledtabular}
\end{table}

\section{Classification of spin singlet pairing states with \texorpdfstring{$C_n$}{Lg} and TRS}
In this section, we classify the possible spin-singlet pairing states constrained by $C_n$ about $z$-axis and TRS. And we also discuss the spontaneous time-reversal symmetry breaking pairings and the induced orbital polarized density-of-states.

\subsection{Classification of pairings}
The pairing potential $\hat{\Delta}(\mathbf{k})$ transforms under the rotation $C_n$ as 
\begin{align}
C_n\hat{\Delta}_J(\mathbf{k})C_n^T=e^{i\frac{2\pi}{n}J}\hat{\Delta}_J(C_n^{-1}\mathbf{k}),
\end{align} 
where $J$ labels the irreducible representations of the $C_n$ point group. For example, $J=0$ is for $A$ representation and $J=2$ is for $B$ representation. 
Firstly, the TRS requires the \textit{coexistence} of $\hat{\Delta}_J$ and $\hat{\Delta}_{-J}$ with equal weight. If the rotation symmetry $C_n$ is further imposed, then $J$ and $-J$ have to be equivalent modulo $n$, i.e. $J\equiv -J \text{ mod }n$. 
The results for the basis functions of $\Psi_s(\mathbf{k})$ and $\mathbf{d}_o(\mathbf{k})$ are summarized in Table~(1) in the main text.
Here, the $k_z$-dependent pairing symmetries are also presented for completeness. However, such pairings are neglected in the main text where we mainly focus on 2D systems. 

When inversion symmetry is also present, it leads to the following constraints for different orbital basis,
\begin{itemize}
\item [1.)] If the inversion symmetry is $\mathcal{I}=\tau_0\sigma_0$, two atomic orbitals have the same parity, 
              it require that $ d_o^2 = 0$. 
\item [2.)] If the inversion symmetry is  $\mathcal{I}=\tau_3\sigma_0$, two atomic orbitals have opposite parities, 
			it require that $ d_o^1 = 0$. 
\item [3.)] If the inversion symmetry is $\mathcal{I}=\tau_1\sigma_0$, two orbitals are the valley indexes, 
			 it require that $ d_o^3 = 0$. 
\end{itemize}

\subsection{Spontaneous TRS-breaking orbital-polarization}
Next, we study spontaneous TRS-breaking and its consequences for a two-band SC with $\{d_{xz},d_{yz}\}$-orbitals. 
$\mathcal{I}=\tau_0\sigma_0$ constrains the orbital $\mathbf{d}_o$-vector to be $(d_1\Psi_o^1(\mathbf{k}),0,d_3\Psi_o^3(\mathbf{k}))$ for Eq.~\eqref{eq-J-2-pairing}.
Under $C_{n}$ ($\mathcal{T}$), the orbital $\mathbf{d}_o$ transforms as $\mathbf{d}_o\to e^{i2\pi J/n}\mathbf{d}_o$ ($\mathbf{d}_o\to \mathbf{d}_o^\ast$). 
Choosing $\sqrt{|d_1|^2+|d_3|^2}=1$, the set of superconducting order parameters are given by $\{\Delta_s,\Delta_o,\mathbf{d}\triangleq(d_1,0,d_3)\}$. 
Furthermore, the orbital orderings can be characterized by $\mathbf{M}_o \propto \sum_{\mathbf{k},\sigma} \langle c^\dagger_{\sigma a}(\mathbf{k}) \bm{\tau}_{ab} c_{\sigma b}(\mathbf{k}) \rangle$.
The total GL free energy preserving the $U(1)\times \mathcal{T} \times C_n \times \mathcal{I}$ symmetries is constructed as,
\begin{align} 
	\mathcal{F}[\Delta_s,\Delta_o,\mathbf{d},\mathbf{M}_o] = \mathcal{F}_1 +\mathcal{F}_2 + \mathcal{F}_{3} + \mathcal{F}_{4},
\end{align}
where 
$\mathcal{F}_1=\frac{1}{2}\alpha_1(T) \vert \Delta_o\vert^2 + \frac{1}{4}\beta_1\vert\Delta_o\vert^4 +\frac{1}{2}\alpha_2(T) \vert \Delta_s\vert^2 + \frac{1}{4}\beta_2\vert\Delta_s\vert^4 + \frac{1}{2}\alpha_M|\mathbf{M}_o|^2$, determining the gap strengths by using $\alpha_{1,2}(T)=\alpha_0^{1,2}(T/T_c^{1,2}-1)$ and $\alpha_0^{1,2},\alpha_M,\beta_{1,2}>0$. 
The second term is 
$\mathcal{F}_2=\beta|d_1|^4+\beta'|d_3|^4$ with  $\beta,\beta'>0$, which determines $\mathbf{d}$.
In addition, there are two possibilities to achieve the spontaneously TRS-breaking states, which are described respectively by $\mathcal{F}_{3}$ and $\mathcal{F}_{4}$,
\begin{align}
	\label{eq-gl-free-energy-f3}
	\mathcal{F}_{3} &=b_1\Delta_s^\ast\Delta_o + b_2(\Delta_s^\ast)^2(\Delta_o)^2  + \text{h.c.}, \\
	\mathcal{F}_{4} &= \gamma_0(T) (\mathbf{d}\times \mathbf{d}^\ast)^2 + i\gamma_1\mathbf{M}_o\cdot(\mathbf{d}\times \mathbf{d}^\ast) + \text{h.c.}.
	\label{eq-gl-free-energy-f4}
\end{align}
The $\mathcal{F}_{3}$ term in Eq.~\eqref{eq-gl-free-energy-f3} helps to develop a relative phase difference between $\Delta_s$ and $\Delta_o$ of being $\pm\pi/2$ when $b_1=0$ and $b_2>0$ \cite{wang_prl_2017}.
As for the $b_2<0$ case, the TRS-breaking is caused solely by the $\mathcal{F}_{4}$ term in Eq.~\eqref{eq-gl-free-energy-f4}. 
For example, $\gamma_0(T)=\gamma_0(T/T_{c}'-1)$ and $T_c'<T_c$, where $T_c'$ is the critical temperature for the spontaneous TRS-breaking inside the superconducting states.
When $T<T_c'$, the orbital $\mathbf{d}_o$-vector becomes complex, then it generates the orbital orderings as
$\mathbf{M}_o=-i\gamma_1/\alpha_M(\mathbf{d}\times \mathbf{d}^\ast)$,
of which only the $y$-component breaks TRS, as illustrated in the main text (see Fig.~1).
More precisely, ${M}_o^y \propto \sum_{\mathbf{k},\sigma} \langle \hat{n}_{\sigma,+}  (\mathbf{k}) - \hat{n}_{\sigma,-}(\mathbf{k}) \rangle$. Here we define $\vert \pm\rangle= \vert 1\rangle + i \vert 2\rangle$ for complex orbitals, thus ${M}_o^y\neq0$ indicates the TRS-breaking orbital-polarization (OP), similar to the time-reversal-odd polarization of the Cooper pairs discussed in Ref.~\cite{brydon_prb_2018,brydon_prx_2019}.

We next solve the Bogoliubov-de-Gennes Hamiltonian, 
\begin{align}
\mathcal{H}_{BdG}\vert E_n(\mathbf{k})\rangle &= E_n(\mathbf{k}) \vert E_n(\mathbf{k})\rangle, \\
\vert E_n(\mathbf{k})\rangle &= (u_{d_{xz},\uparrow}^n,u_{d_{xz},\downarrow}^n,v_{d_{xz},\uparrow}^n,v_{d_{xz},\downarrow}^n,u_{d_{yz},\uparrow}^n,
u_{d_{yz},\downarrow}^n,   v_{d_{yz},\uparrow}^n, v_{d_{yz},\downarrow}^n)^T,
\end{align}
The quasi-particle spectrum is plotted in the main text (see Fig.~1), where two distinct gaps appear.
Then, we calculate the atomic orbital-polarized density of states (DOS),
\begin{align}
D_\kappa(E) = \frac{1}{2}\sum_{\sigma,n,\mathbf{k}} |u_{\kappa,\sigma}^n|^2 \delta\left(E - E_n(\mathbf{k}) \right),
\end{align} 
where $u_{\kappa,\sigma}^n = u^n_{d_{xz},\sigma}-i \kappa u^n_{d_{yz},\sigma}$ with $\kappa=\pm$ for $d_{xz}\pm i d_{yz}$ orbitals, and $\delta(x)$ is the delta function. 
In Fig.~\ref{figadd0}, the numerical results helps to confirm a two-gap feature due to the spontaneous breaking of TRS, compared with the quasi-particle spectrum. 
Moreover, $D_+\neq D_-$ at finite energy clearly indicates that the DOS is orbital-polarized, which is consistent with the GL analysis.
The orbital-spin conversion would lead to the spin-polarized DOS~\cite{hu_arxiv_xxx}.

\begin{figure}[!htbp]
	\centering
	\includegraphics[width=0.7\linewidth]{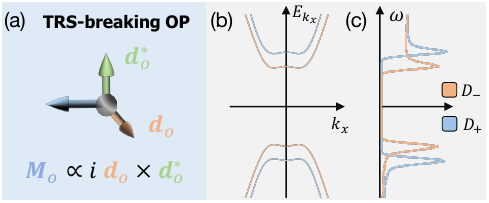}
	\caption{The TRS-breaking effects for spin-singlet two-band SCs. 
		(a) A complex orbital $\mathbf{d}_o$-vector spontaneously breaks TRS and may give rise to the TRS-breaking OP with $\mathbf{M}_o \propto i\mathbf{d}_o\times\mathbf{d}_o^\ast$, which is the same plot in Fig.~(1b) in the main text. 
		(b) The quasi-particle spectrum along $\mathbf{k}=(k_x,0)$ is shown. 
		(c) The orbital-polarized DOS $D_{\pm}$ with $\pm$ representing $d_{xz}\pm i d_{yz}$ are exhibited. 
		Parameters used are: $\lambda_o=0.1$, $\Delta_o=1$, $\Delta_s=0.1$, $\mu=-0.5$, $t_0=1$, $\mathbf{d}_o=(\cos\tfrac{\pi}{20},0,i\sin\tfrac{\pi}{20})$.
		\label{figadd0}
	}
\end{figure}

\section{The stability of orbital $\mathbf{d}_o$-vector under crystal fields}
In the Method section of the main text, we derived our main result up to first-order in the coupling $\lambda_o$. Here, we first address the situation where multiple pairing channels with possibly different pairing strengths coexist. Then we show the second-order result in $\lambda_o$.

\subsection{First-order result applied to multiple coexisting pairing channels}
We have the following first-order result,
\begin{align}\label{sm-tc-lambda0}
\ln\left(\frac{T_c}{T_{c0}} \right) =\int_{S} \frac{d\Omega}{2\pi} \, \mathcal{C}_0(T_{c}) \left( \left\vert \mathbf{d}_o \right\vert^2 -  \left\vert \mathbf{d}_o\cdot\hat{\mathbf{g}}_o \right\vert^2 \right),
\end{align}
which is derived with the assumption that there is only one pairing channel. Here, we elaborate on a subtlety mentioned in the main text that might arise due to coexisting multiple pairing channels belonging to different 1d irreducible representations. In the weak-coupling theory and without orbital hybridization, the critical temperature for a particular channel $\Gamma$ is simply obtained by solving the linearized gap equation $v^\Gamma \chi_0(T_{c0})=1$ and the solution is given by $T_{c0}^\Gamma=\frac{2e^{\gamma}\omega_D}{\pi k_B}e^{-\frac{1}{v^\Gamma N_0}}$. This means that the critical temperature in each channel is solely determined by the strength of the pairing interaction in that particular channel. The leading instability channel has the largest pairing interaction, which determines the $T_c$. However, when orbital hybridization is considered, the story could change. Depending on the relation between $\mathbf{d}_o^\Gamma$ and $\mathbf{g}_o$, some pairing channels will be suppressed more than the others. Therefore, the previous leading instability channel could become sub-leading in the presence of orbital hybridization. In Figure~\ref{fig:sm-fig-add} we take the $J=0$ and $J=2$ representations under $C_4$ as an example, with the assumption that $v^{J=0}>v^{J=2}$. We see that $T_c^{J=0}$ starts higher than $T_c^{J=2}$, but since $\mathbf{d}_o^{J=2}$ is parallel $\mathbf{g}_o$ and $\mathbf{d}_o^{J=0}$ is not, $T_c^{J=2}$ is not suppressed by $\lambda_o$ whereas $T_c^{J=0}$ is suppressed and eventually becomes lower than $T_c^{J=2}$.
\begin{figure}
    \centering
    \includegraphics[width=0.5\textwidth]{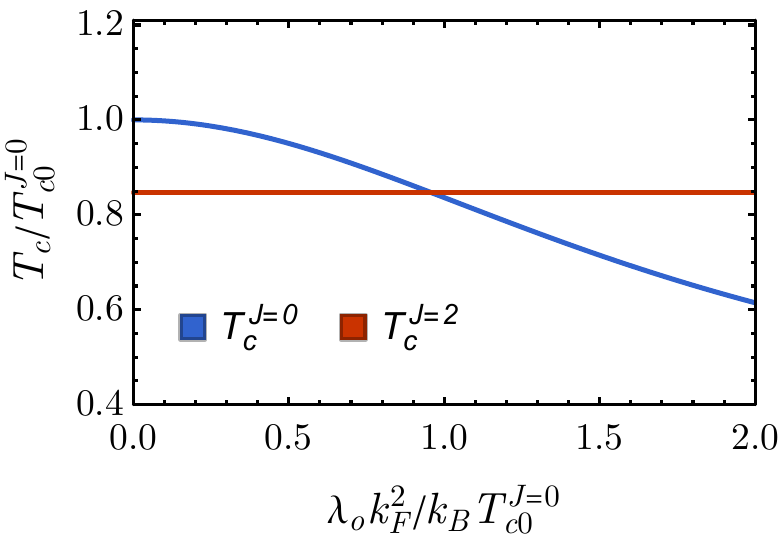}
    \caption{Different behaviors of $T_c^{J=0}$ and $T_c^{J=2}$ under orbital hybridization. Here $\mathbf{d}_o^{J=0}=k_F^{-2}(k_x^2-k_y^2,0,2k_xk_y),\mathbf{d}_o^{J=2}=\frac{1}{\sqrt{2}}(1,0,1)$, and $\mathbf{g}_o=k_F^2(1,0,1)$.}
    \label{fig:sm-fig-add}
\end{figure}

\subsection{Second-order approximated results}
In this subsection, we consider the coupling between orbital-independent pairing ($\Psi_s(\mathbf{k})$-part in Eq.~\eqref{sm-eq-singlet-pairing-tot}) and orbital-dependent pairing ($\mathbf{d}_o(\mathbf{k})$-part in Eq.~\eqref{sm-eq-singlet-pairing-tot}), and study the second-order approximated results for the above conclusion.

The attractive interaction is now decomposed as 
\begin{equation}
    V^{s_1a,s_2b}_{s_1'a',s_2'b'}(\mathbf{k},\mathbf{k}') = -v_0 \lbrack\mathbf{d}_o(\mathbf{k})\cdot\bm{\tau}i\sigma_2 \rbrack_{s_1a,s_2b}  \lbrack\mathbf{d}_o(\mathbf{k}')\cdot\bm{\tau}i\sigma_2 \rbrack_{s_1'a',s_2'b'}-v_1 \lbrack\Psi_s(\mathbf{k})i\sigma_2 \rbrack_{s_1,s_2}  \lbrack\Psi_s(\mathbf{k}')i\sigma_2 \rbrack_{s_1',s_2'},
\end{equation}
where $v_0$ is the interaction strength in the orbital-dependent channel and $v_1$ is the interaction strength in the orbital-independent channel. 
And they belong to the same representation of symmetry groups, leading to the coupled linearized gap equation, 
\begin{equation}
\text{Det}\begin{pmatrix}v_0\chi(T)-1 & v_0\chi_{os}(T)\\v_1\chi_{so}(T) & v_1\chi_s(T)-1\end{pmatrix}=0,
\end{equation}
where
\begin{equation}
\begin{split}
    &\chi_{os}(T)\equiv -\frac{1}{\beta}\sum_{\mathbf{k},\omega_n} \text{Tr}\left\lbrack (\mathbf{d}_o(\mathbf{k})\cdot\bm{\tau}i\sigma_2)^\dagger G_e(\mathbf{k},i\omega_n) (\Psi_s(\mathbf{k})i\sigma_2) G_h(-\mathbf{k},i\omega_n) \right\rbrack,\\
    &\chi_{so}(T)\equiv -\frac{1}{\beta}\sum_{\mathbf{k},\omega_n} \text{Tr}\left\lbrack (\Psi_s(\mathbf{k})i\sigma_2)^\dagger G_e(\mathbf{k},i\omega_n) (\mathbf{d}_o(\mathbf{k})\cdot\bm{\tau}i\sigma_2) G_h(-\mathbf{k},i\omega_n) \right\rbrack.
\end{split}
\end{equation}
It leads to 
\begin{align}
(v_0\chi(T)-1) (v_1\chi_s(T)-1)  - v_0v_1\chi_{so}(T)\chi_{os}(T) = 0,
\end{align}
Considering the $v_0>v_1$ case firstly, then, the bare $T_c$ of orbital-dependent pairings are larger than that of orbital-independent pairings, we have
\begin{align}\label{sm-eq-chi0-tot-do-part}
v_0\chi(T)-1 - \frac{v_0v_1\chi_{so}(T)\chi_{os}(T)}{v_1\chi_s(T)-1}=0,
\end{align}
from which, we define the total superconductivity susceptibility as,
\begin{align}\label{sm-eq-chi0-tot}
\chi'(T) = \chi(T) + \frac{\chi_{so}(T)\chi_{os}(T)}{1/v_1 - \chi_s(T)}
\end{align}
where $ \chi(T)$ has been calculated in the above subsection (see Eq.~\eqref{sm-eq-chi0-do}), and the second part is the second-order correction.
After tracing out the spin degrees of freedom, we have $\chi_{os}(T)=\chi_{so}(T)$. Following the same procedure as in the first-order case, we have 
\begin{equation}
\begin{split}
   \chi_{os}(T)&= -\frac{2N_0}{\beta} \int_{-\omega_D}^{+\omega_D}d\epsilon\int_{S} \frac{d\Omega}{2\pi} \sum_{\omega_n}\sum_{\alpha} \alpha G_e^{\alpha}(\mathbf{k},i\omega_n) G_h^{\alpha}(\mathbf{k},i\omega_n)\lbrack (\mathbf{d}_o(\mathbf{k})\cdot\hat{\mathbf{g}}_o(\mathbf{k}))\Psi_s(\mathbf{k}) \rbrack\\
   &=-\frac{2N_0}{\beta} \int_{-\omega_D}^{+\omega_D}d\epsilon\int_{S} \frac{d\Omega}{2\pi} \sum_{\omega_n}\lbrack G_e^{+}(\mathbf{k},i\omega_n) G_h^{+}(\mathbf{k},i\omega_n)-G_e^{-}(\mathbf{k},i\omega_n) G_h^{-}(\mathbf{k},i\omega_n)\rbrack\lbrack (\mathbf{d}_o(\mathbf{k})\cdot\hat{\mathbf{g}}_o(\mathbf{k}))\Psi_s(\mathbf{k}) \rbrack,
   \end{split}
\end{equation}
which would vanish if $\lambda_o=0$, i.e. no orbital hybridization, based on the definitions of $G_{e/h}^{+/-}$, which in turn, reproduces the first-order calculation above. At non-zero, but small $\lambda_o$ ($\lambda_o k_F^2<\mu$), $\chi_{os}(T)$ will also be small but non-zero. For convenience of discussion, we define
\begin{equation}
    \delta(T,\lambda_o)\equiv -\frac{N_0}{\beta} \int_{-\omega_D}^{+\omega_D}d\epsilon \sum_{\omega_n}\lbrack G_e^{+}(\mathbf{k},i\omega_n) G_h^{+}(\mathbf{k},i\omega_n)-G_e^{-}(\mathbf{k},i\omega_n) G_h^{-}(\mathbf{k},i\omega_n)\rbrack,
\end{equation}
where $\delta(T,\lambda_o)\sim \lambda_ok_F^2/\mu$ would vanish at leading order (see Eq.~\eqref{sm-eq-first-order-GeGh}). Then we have 
\begin{equation}
    \chi_{os}(T)=2\delta(T,\lambda_o)\int_{S} \frac{d\Omega}{2\pi} \lbrack (\mathbf{d}_o(\mathbf{k})\cdot\hat{\mathbf{g}}_o(\mathbf{k}))\Psi_s(\mathbf{k}) \rbrack.
\end{equation}
With this, the total superconductivity susceptibility in Eq.~\eqref{sm-eq-chi0-tot} becomes
\begin{equation}
    \chi'(T)=\chi(T)+\frac{\chi_{os}^2(T)}{2N_0\log(T/T_s)} \equiv \chi(T)+\delta\chi(T).
\end{equation}
where $\delta\chi(T)$ is the second-order correction due to the coupling between orbital-independent pairings ($\Psi_s(\mathbf{k})$-part in Eq.~\eqref{sm-eq-singlet-pairing-tot}) and orbital-dependent pairing ($\mathbf{d}_o(\mathbf{k})$-part in Eq.~\eqref{sm-eq-singlet-pairing-tot}),
\begin{align}\label{sm-eq-delta-chi-correction}
\delta\chi(T) = \frac{2\delta^2(T,\lambda_o)\left(\int_{S} \frac{d\Omega}{2\pi} \lbrack (\mathbf{d}_o(\mathbf{k})\cdot\hat{\mathbf{g}}_o(\mathbf{k}))\Psi_s(\mathbf{k}) \rbrack\right)^2}{N_0\log(T/T_s)}.
\end{align}
Since we assumed $v_0>v_1$, i.e. $T_{c0}>T_s$, then the actual transition temperature would be $T_c\sim T_{c0}>T_s$, giving $\log(T_c/T_s)>0$. As a result, the correction to the susceptibility is positive: $\delta\chi(T)>0$.
Then following the same procedure as in the previous section, we have 
\begin{align}\label{sm-tc-lambda0}
\ln\left(\frac{T_c}{T_{c0}} \right) =\int_{S} \frac{d\Omega}{2\pi} \, \mathcal{C}_0(T_{c0}) \left( \left\vert \mathbf{d}_o \right\vert^2 -  \left\vert \mathbf{d}_o\cdot\hat{\mathbf{g}}_o \right\vert^2 \right)+\frac{\delta \chi(T_{c0})}{N_0},
\end{align}
where the first part is the first-order result (see Eq.~\eqref{sm-tc-lambda0} or Eq.~(5) in the main text; order as $\mathcal{O}(\lambda_o k_F^2/\mu)$), and the second part is the second-order result (order as $\mathcal{O}((\lambda_o k_F^2/\mu)^2)$).
Therefore, we conclude that,
\begin{itemize}
\item {\bf The first part}: it determines the direction of orbital $\mathbf{d}_o$-vector to be $\mathbf{d}_o \parallel \hat{\mathbf{g}}_o$. Because of $\mathcal{C}_0(T_{c0}) \le 0$ , once $\mathbf{d}_o \parallel \hat{\mathbf{g}}_o$ at any momentum $\mathbf{k}$, the first part vanishes.
\item {\bf The second part}: it relates the form $\Psi_s(\mathbf{k})$ to the orbital $\mathbf{d}_o$-vector: $\mathbf{d}_o \propto \Psi_s(\mathbf{k}) \hat{\mathbf{g}}_o$. Thus, the second part becomes maximum, leading to the maximal increase of $T_{c0}$.
\end{itemize}
Here we have used the fact,
\begin{align}\label{sm-eq-factor}
\left( \int_{S} \frac{d\Omega}{2\pi} \lbrack \Psi_i(\mathbf{k})  \Psi_j(\mathbf{k}) \rbrack \right)^2 \le 1, \text{ for any scalar }\Psi_i,\;  \int_{S} \frac{d\Omega}{2\pi} \lbrack \Psi_i(\mathbf{k})  \Psi_i(\mathbf{k}) \rbrack =1 .
\end{align}
And $\left( \int_{S} \frac{d\Omega}{2\pi} \lbrack \Psi_i(\mathbf{k})  \Psi_j(\mathbf{k}) \rbrack \right)^2 $ reaches 1 only when $i=j$.

Next, we briefly discuss the case where $v_0<v_1$. 
The same result can be similarly argued. 
In this case, the dominant pairing channel is the orbital-independent pairing ($T_s>T_{c0}$), which can induce the orbital $\mathbf{d}_o$-vector via their couplings. 
Similar to Eq.~\eqref{sm-eq-chi0-tot-do-part}, we define the total superconductivity susceptibility for orbital-independent pairings,
\begin{align}
v_1\chi_s(T) - 1 - \frac{v_0v_1\chi_{so}(T)\chi_{os}(T)}{v_0\chi(T)-1}=0,
\end{align}
which leads to
\begin{align}
\chi_s'(T) = \chi_s(T) + \frac{\chi_{so}(T)\chi_{os}(T)}{1/v_0 - \chi(T)} = N_0\ln\left( \tfrac{2e^\gamma \omega_D}{\pi k_B T} \right) + \frac{\chi_{os}^2(T)}{N_0 \ln (\frac{T}{T_{c0}}) - \int_{S} \frac{d\Omega}{2\pi} \, \mathcal{C}_0(T_{c0}) \left( \left\vert \mathbf{d}_o \right\vert^2 -  \left\vert \mathbf{d}_o\cdot\hat{\mathbf{g}}_o \right\vert^2 \right) },
\end{align}
thus,
\begin{align}\label{sm-eq-ts-tot}
\ln\left(\frac{T_c}{T_s}\right) = 2\delta^2(T_c,\lambda_o) \times
\frac{\left(\int_{S} \frac{d\Omega}{2\pi} \lbrack (\mathbf{d}_o(\mathbf{k})\cdot\hat{\mathbf{g}}_o(\mathbf{k}))\Psi_s(\mathbf{k}) \rbrack\right)^2}{N_0 \ln (\frac{T_c}{T_{c0}}) - \int_{S} \frac{d\Omega}{2\pi} \, \mathcal{C}_0(T_{c0}) \left( \left\vert \mathbf{d}_o \right\vert^2 -  \left\vert \mathbf{d}_o\cdot\hat{\mathbf{g}}_o \right\vert^2 \right) } \ge 0,
\end{align}
here $T_c\sim T_s > T_{c0}$ so that $ \ln (\frac{T_c}{T_{c0}}) >0$.
The correction is in order of $\mathcal{O}((\lambda_o k_F^2/\mu)^2)$.
Therefore, 
\begin{itemize}
\item {\bf The denominate}: it determines the direction of orbital $\mathbf{d}_o$-vector to be $\mathbf{d}_o \parallel \hat{\mathbf{g}}_o$. Because of $\mathcal{C}_0(T_{c0}) \le 0$, then, $- \int_{S} \frac{d\Omega}{2\pi} \, \mathcal{C}_0(T_{c0}) \left( \left\vert \mathbf{d}_o \right\vert^2 -  \left\vert \mathbf{d}_o\cdot\hat{\mathbf{g}}_o \right\vert^2 \right) \ge0$, once $\mathbf{d}_o \parallel \hat{\mathbf{g}}_o$ at any momentum $\mathbf{k}$, the denominate is positive and minimum.
\item {\bf The numerator}: it determines the local magnitude of orbital $\mathbf{d}_o$-vector to be $\mathbf{d}_o \propto \Psi_s(\mathbf{k}) \hat{\mathbf{g}}_o$. Thus, the numerator becomes maximum, leading to the increase of $T_{c0}$ maximally.
\end{itemize}

Therefore, according to both Eq.~\eqref{sm-tc-lambda0} and Eq.~\eqref{sm-eq-ts-tot}, we conclude that the orbital $\mathbf{d}_o$-vector that is parallel with orbital hybridization $\mathbf{g}_o$-vector could be generally stabilized in real materials. And we find that
\begin{align}
\mathbf{d}_o = \pm \Psi_s(\mathbf{k}) \hat{\mathbf{g}}_o ,
\end{align}
which is shown in Eq.~(6) in the main text.
However, it has also a $Z_2$ phase $\pm$, which can be pinned by taking higher order corrections into account.

\section{Formation of the pairing near Fermi surface in band picture}
Here, we provide another perspective on the pairing in orbital channel near the Fermi surface (FS) by looking at the total free energy of the system in band picture, where the pairing amplitude is treated perturbatively. 

In the presence of orbital hybridization or nematic order, the Hamiltonian without pairing is given by 
\begin{equation}
\mathcal{H}_0=(\epsilon_{\mathbf{k}}-\mu)\tau_0\sigma_0+\lambda (\mathbf{g}\cdot\bm{\tau})\sigma_0,
\end{equation}
where $\lambda$ is taken to be positive. The degeneracy in the orbital channel will be lifted, whereas the spin channel still has the double-degeneracy. Effectively, the vector $\mathbf{g}$ acts as a ``Zeeman field'' in the orbital space, and the pseudo-spin will be parallel or anti-parallel to the field for the two splitting levels. And we notice that
\begin{align}
[\mathbf{g}\cdot\bm{\tau}, \mathcal{H}_0] = 0.
\end{align}
More precisely, the two eigenstates of $ \mathcal{H}_0$ can be denoted by $|E_{\pm}\rangle\equiv |\hat{\mathbf{g}};\pm\rangle$, where $+/-$ means parallel/anti-parallel (eigenvalues of the symmetry operator $\hat{\mathbf{g}}\cdot\bm{\tau}$). 
Please note that $\hat{\mathbf{g}} = \mathbf{g}/\vert \mathbf{g} \vert$. And 
\begin{align}
\mathcal{H}_0|E_{\pm}\rangle=E_{\pm}|E_{\pm}\rangle, \text{ with } E_\pm = \epsilon_{\mathbf{k}}-\mu \pm \lambda \vert \mathbf{g}(\mathbf{k}) \vert.
\end{align}
with $\lambda>0$.
Setting $E_\pm = 0$, it gives rise to two FSs (labeled as FS$_\pm$) with energy splitting as $2\lambda \vert \mathbf{g}(\mathbf{k}) \vert$, which is approximately as $\sim \lambda k_F^2$ with respect to $\epsilon_{k_F}=\mu$. 

Now we consider the spin-singlet pairing part in the original basis,
\begin{equation}\label{sm-eq-sc-do}
\mathcal{H}_{\Delta}= \sum_{\mathbf{k}} (c_{1,\uparrow}^\dagger(\mathbf{k}) , c_{2,\uparrow}^\dagger (\mathbf{k}))  
\left\lbrack \Delta_s\Psi(\mathbf{k})\tau_0 +  \Delta_o(\mathbf{d}_o(\mathbf{k})\cdot\bm{\tau}) \right\rbrack 
(c_{1,\downarrow}^\dagger(-\mathbf{k}) , c_{2,\downarrow}^\dagger (-\mathbf{k})) ^T +\text{H.c.}.
\end{equation}
The BdG Hamiltonian is then given by
\begin{equation}
\mathcal{H}_{BdG}=\left((\epsilon_{\mathbf{k}}-\mu)\tau_0\sigma_0+\lambda (\mathbf{g}\cdot\bm{\tau})\sigma_0\right)\gamma_3+\left\lbrack \Delta_s\Psi(\mathbf{k}) +  \Delta_o(\mathbf{d}_o(\mathbf{k})\cdot\bm{\tau}) \right\rbrack(i\sigma_2)\gamma_2,
\end{equation}
where $\gamma_i$ are the Pauli matrices in particle-hole channel. 

Next, we consider weak-coupling limit (infinitesimal pairing strength, namely, $\Delta_o\to 0$) and we use the band picture to study the pairing Hamiltonian. 
For this purpose, we rewrite $\mathcal{H}_0$ as
\begin{align}
\mathcal{H}_0 = \sum_{\mathbf{k},\tau,s} \;  E_{\tau}(\mathbf{k}) c_{\tau,s}^\dagger(\mathbf{k})   c_{\tau,s}(\mathbf{k}) .
\end{align}
where $\tau=\pm$ is the band index and $s$ is for spin.
The unitary transformation matrix $U(\mathbf{k})$ in the orbital subspace leads to the diagonalization of $\mathcal{H}_0$,
\begin{align}
 U^\dagger(\mathbf{k}) \lbrack (\epsilon_{\mathbf{k}}-\mu)\tau_0+\lambda (\mathbf{g}\cdot\bm{\tau}) \rbrack U(\mathbf{k}) = \text{Diag}[E_+(\mathbf{k}), E_-(\mathbf{k})],
\end{align}
where spin index has been dropped and the 2-by-2 $U(\mathbf{k})$ can be expressed by the eigenstates of $\mathcal{H}_0$,
\begin{align}
 U(\mathbf{k})  = \left\{\vert E_+(\mathbf{k}) \rangle , \vert E_-(\mathbf{k}) \rangle \right\} .
\end{align}
Thus, $U^\dagger(\mathbf{k})  U(\mathbf{k})   = 1$.
And the time-reversal symmetry requires that
\begin{align}
 U(\mathbf{k}) = \left( U(-\mathbf{k})\right)^\ast.
\end{align}
Acting on the basis, we have
\begin{align}
(c_{1,s}^\dagger(\mathbf{k}), c_{2,s}^\dagger(\mathbf{k})) = (c_{+,s}^\dagger(\mathbf{k}) , c_{-,s}^\dagger (\mathbf{k})) U^\dagger(\mathbf{k}) 
\end{align}
where $s$ is for spin. We then project the spin-singlet pairing Hamiltonian in Eq.~\eqref{sm-eq-sc-do} into the band basis, thus, the spin-singlet pairing Hamiltonian becomes
\begin{align}   
\mathcal{H}_{\Delta} = \sum_{\mathbf{k}} (c_{+,\uparrow}^\dagger(\mathbf{k}) , c_{-,\uparrow}^\dagger (\mathbf{k})) 
\left\{ U^\dagger(\mathbf{k})  \left\lbrack \Delta_s\Psi(\mathbf{k}) +  \Delta_o(\mathbf{d}_o(\mathbf{k})\cdot\bm{\tau}) \right\rbrack U(\mathbf{k}) \right\} (c_{+,\downarrow}^\dagger(-\mathbf{k}) , c_{-,\downarrow}^\dagger (-\mathbf{k}))^T.
\end{align}
Here  $U(\mathbf{k}) = \left( U(-\mathbf{k})\right)^\ast$ has been used.
Therefore, the spin-singlet pairing potential in the band basis becomes 
\begin{align}
\Delta_{\text{band}}(\mathbf{k})  =  U^\dagger(\mathbf{k})  \left\lbrack \Delta_s\Psi(\mathbf{k})\tau_0 +  \Delta_o(\mathbf{d}_o(\mathbf{k})\cdot\bm{\tau}) \right\rbrack U(\mathbf{k}).
\end{align}
Thus, the orbital-independent pairings only lead to the intra-band pairings, while the orbital-dependent pairings can give rise to both intra-band and inter-band pairings.

First of all, we focus on the pure orbital $\mathbf{d}_o$-vector (orbital-dependent pairings) by assuming $\Delta_s=0$.
To separate the intra-band pairings from the inter-band pairings,
we decompose the orbital $\mathbf{d}_o$-vector as
\begin{align}
\mathbf{d}_o(\mathbf{k}) = d_{\parallel}(\mathbf{k})  \hat{\mathbf{g}}(\mathbf{k}) +\mathbf{d}_{\perp}(\mathbf{k}),
\end{align}
where $d_{\parallel}(\mathbf{k})= \mathbf{d}_o(\mathbf{k}) \cdot \hat{\mathbf{g}}(\mathbf{k})  $ and $\mathbf{d}_{\perp}(\mathbf{k})\cdot \hat{\mathbf{g}}(\mathbf{k})  = 0$. Thus,
\begin{itemize}
\item {\bf The parallel component}: the $d_{\parallel}(\mathbf{k})  \hat{\mathbf{g}}(\mathbf{k}) $-part commutes with $\mathbf{g}\cdot\bm{\tau}$, hence only generates intra-band pairing for the two FSs. Therefore, the projected intra-band pairing Hamiltonian reads,
\begin{align}
\mathcal{H}_{\text{intra-band},\Delta} = \sum_{\mathbf{k},\tau} \left(\tau \Delta_o d_{\parallel}(\mathbf{k}) \right)  
\left\lbrack  c_{\tau,\uparrow}(\mathbf{k})   c_{\tau,\downarrow}(-\mathbf{k}) -  c_{\tau,\downarrow}(\mathbf{k})   c_{\tau,\uparrow}(-\mathbf{k})   \right\rbrack + \text{H.c.}.
\end{align}
Here $\tau=\pm$ indicates that the intra-band pairing strengths on two FSs are opposite. 
And $d_{\parallel}(\mathbf{k})  = d_{\parallel}(-\mathbf{k}) $, the intra-band pairing is a even-parity pairing.
\item {\bf The perpendicular component}:  the $\mathbf{d}_{\perp}(\mathbf{k})$-part would mix the two states $|E_{\pm}\rangle$, then it only produces inter-band pairings. 
At a fixed $\mathbf{k}$, we now perform a rotation,
\begin{align}
O(3) \text{ rotation: } R_o \hat{\mathbf{g}}_o R_o^\dagger = (0,0, g_z')  \text{ and } R_o \mathbf{d}_{\perp} R_o^\dagger = (g_x',g_y',0)
\end{align}
at the same time, we perform a rotation in the orbital subspace
\begin{align}
SU(2) \text{ rotation: } R_\tau \tau_{x,y,z} R_\tau^\dagger = \tau'_{x,y,z}
\end{align}
Due to this rotation, the perpendicular components only couple $\vert E_+\rangle $ with $\vert E_-\rangle$. This proves that
\begin{align}
\langle E_\tau \vert \mathbf{d}_{\perp}\cdot\bm{\tau} \vert E_\tau \rangle =0.
\end{align}
Here $\vert E_\tau\rangle$ are eigenstates of $\hat{\mathbf{g}}_o\cdot\bm{\tau}$.
Thus, the projected inter-band pairing Hamiltonian reads,
\begin{align}
\mathcal{H}_{\text{inter-band},\Delta} = \sum_{\mathbf{k},\tau} \Delta_{\tau,-\tau}(\mathbf{k})
\left\lbrack  c_{\tau,\uparrow}(\mathbf{k})   c_{-\tau,\downarrow}(-\mathbf{k}) -  c_{\tau,\downarrow}(\mathbf{k})   c_{-\tau,\uparrow}(-\mathbf{k})   \right\rbrack + \text{H.c.}.
\end{align}
where $\Delta_{\tau,-\tau}(\mathbf{k})=\Delta_o \langle E_{\tau}(\mathbf{k}) \vert \mathbf{d}_{\perp}(\mathbf{k})\cdot\bm{\tau} \vert E_{-\tau}(\mathbf{k}) \rangle  $.
In the limit $\lambda k_F^2 \gg \Delta_o$ (i.e., band splitting is much larger than the pairing gap), the inter-band pairing is not energetically favorable in the weak-coupling pairing theory (i.e., attractive interaction is infinitesimal small). 
It means the inter-band pairing will be severely suppressed if we increase the orbital hybridization $\lambda$, consistent with the calculation in the main text (see Fig.~(2)).
\end{itemize}
From the above analysis, we conclude that the orbital $\mathbf{d}_o$-vector should be parallel with the orbital hybridization $\mathbf{g}$.

To determine the magnitude of the orbital $\mathbf{d}_o$-vector, we turn on the orbital-independent pairing $\Delta_s\neq0$.
We also assume both pairing channels are small. 
Assuming the two FSs have DOS $N_{\pm}$ near the FS (ignoring the momentum-dependence if the FS is almost isotropic), then the total condensation energy per volume and per spin of the two intra-band pairings is given by
\begin{equation}
\delta E  =  N_+\sum_{\mathbf{k}\in \text{FS}_+} \left(\Delta_s\Psi_s(\mathbf{k}) +\Delta_od_{\parallel}(\mathbf{k}) \right)^2+N_-\sum_{\mathbf{k}\in \text{FS}_-}\left(\Delta_s\Psi_s(\mathbf{k}) -\Delta_od_{\parallel}(\mathbf{k}) \right)^2,
\end{equation}
with the approximation $\lambda k_F^2 \ll \mu$, $\delta E$ becomes
\begin{align}
\delta E =(N_++N_-) \sum_{\mathbf{k} \in \text{FS}}  [\Delta_s^2(\Psi_s(\mathbf{k}))^2+\Delta_o^2(d_{\parallel}(\mathbf{k}) )^2]+2(N_+-N_-)\Delta_s\Delta_o \sum_{\mathbf{k}\in \text{FS}} (\Psi_s(\mathbf{k})d_{\parallel}(\mathbf{k})),
\end{align}
where FS is for $\lambda\to0$. And we see that in order to maximize the condensation energy, we require
\begin{align}
d_{\parallel}(\mathbf{k}) &\propto \Psi_s(\mathbf{k}), \text{ if } \text{sign}[(N_+-N_-)\Delta_s\Delta_o] = 1, \\
d_{\parallel}(\mathbf{k}) &\propto -\Psi_s(\mathbf{k}), \text{ if } \text{sign}[(N_+-N_-)\Delta_s\Delta_o] = -1, 
\end{align}
according to Eq.~\eqref{sm-eq-factor}.
The results from the weak-coupling limit are the same as the calculations from linearized gap equations.

\section{The coexistence of orbital $\mathbf{d}_o$-vector and nematic order}
In this section, we discuss the interaction effects on the stability of orbital $\mathbf{d}_o$-vector for a two-band superconductor.
A general density-density interaction, including both inter-band and intra-band terms, is,
\begin{align}
\mathcal{H}_{int} = v_1(\hat{n}_{1\uparrow}+\hat{n}_{1\downarrow})(\hat{n}_{2\uparrow}+\hat{n}_{2\downarrow})+ v_2(\hat{n}_{1\uparrow}\hat{n}_{1\downarrow}+\hat{n}_{2\uparrow}\hat{n}_{2\downarrow}),
\end{align}
where $\hat{n}$ is the electron density operator and $v_{1,2}$ are interaction strengths. 
With the following mean-field decomposition, we can define the nematic ordering that spontaneously breaks the rotational symmetry. 
\begin{align}
\mathcal{H}_{MF} =\hat{n}_{1\uparrow}\left(\frac{v_1}{2}\langle \hat{n}_2\rangle+v_2\langle \hat{n}_{1\downarrow}\rangle\right)
+\hat{n}_{1\downarrow}\left(\frac{v_1}{2}\langle \hat{n}_2\rangle+v_2\langle \hat{n}_{1\uparrow}\rangle\right)
+\hat{n}_{2\uparrow}\left(\frac{v_1}{2}\langle \hat{n}_1\rangle+v_2\langle \hat{n}_{2\downarrow}\rangle\right)
+\hat{n}_{2\downarrow}\left(\frac{v_1}{2}\langle \hat{n}_1\rangle+v_2\langle \hat{n}_{2\uparrow}\rangle\right).
\end{align}
For the purpose of our discussion, we assume there is no spin ferromagnetism, i.e. $\langle \hat{n}_{\alpha \uparrow}\rangle=\langle \hat{n}_{\alpha \downarrow}\rangle=\frac{1}{2}\langle \hat{n}_{\alpha }\rangle$, then the mean field Hamiltonian simplifies to
\begin{equation}
\begin{split}
\mathcal{H}_{MF}&=\hat{n}_{1}\left(\frac{v_1}{2}\langle \hat{n}_2\rangle+\frac{v_2}{2}\langle \hat{n}_{1}\rangle\right)+\hat{n}_{2}\left(\frac{v_1}{2}\langle \hat{n}_1\rangle+\frac{v_2}{2}\langle \hat{n}_{2}\rangle\right)
\equiv \hat{n}_1\Phi_1+\hat{n}_2\Phi_2.
\end{split}
\end{equation}
Then the nematic order parameter can be defined as 
\begin{equation}
\Phi\equiv\Phi_1-\Phi_2=\frac{1}{2}\left(v_2-v_1\right)\left(\langle\hat{n}_1\rangle-\langle\hat{n}_2\rangle\right).
\end{equation}
Now the stage is set to define the total orbital hybridization as,
\begin{align}
\mathbf{g}_{tot} = \mathbf{g}_o + t_{nem} \mathbf{g}_{nem}
\end{align}
where $\mathbf{g}_{nem}=(0,0,\Phi)$ and $t_{nem}=\lambda_{nem}/\lambda_o$.
Then, we apply Eq.~\eqref{sm-tc-lambda0} to study the stability of orbital $\mathbf{d}_o$-vector when nematic order develops above superconducting $T_{c0}$. The results are summarized in Figure~\ref{sm-fig1}.

\begin{figure}[!htbp]
	\centering
	\includegraphics[width=0.95\linewidth]{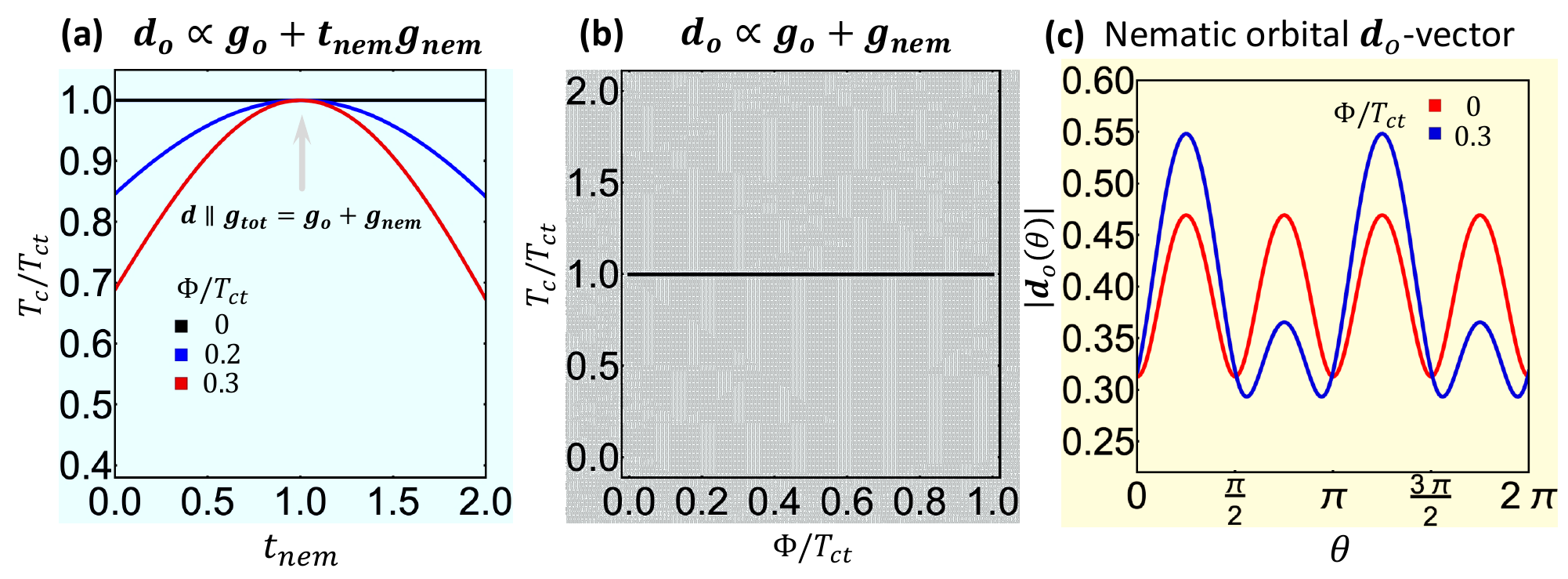}  
	\caption{The coexistence of orbital $\mathbf{d}_o$-vector and nematic order. In (a), each $t_{nem}$ corresponds to a particular form of the $\mathbf{d}$-vector, which determines $T_c$ based on Eq.~\eqref{sm-tc-lambda0}. The three curves correspond to three different values for the nematic order $\Phi$. The $T_c$ is not suppressed by the nematic order as long as $\mathbf{d}_o\parallel \mathbf{g}_{tot}$, i.e. $t_{nem}=1$. The $t_{nem}=1$ case is further illustrated in (b), where it is shown that the magnitude of the nematic order does not change $T_c$ (up to the order of approximation made in Eq.~\eqref{sm-tc-lambda0}). (c) shows non-zero nematic order breaks the original $C_4$ (red line) down to $C_2$ (blue line). Here $\mathbf{g}_o=(k_x^2-k_y^2,0,3k_xk_y)$. }
	\label{sm-fig1}
\end{figure}

\section{Application to single-layer graphene superconductor}
Based on the discussion in the main text, the nematic d-vector is characterized by $\mathbf{g}_{tot}=(g_{int,1},0,0)$, with
\begin{align}
g_{int,1} = 1 + 2t_1 k_x k_y +  t_2(k_x^2 - k_y^2).
\end{align}
A closed Fermi surface (FS) can be parametrized by $k_F(\theta)$. 
By using Eq.~\eqref{sm-tc-lambda0}, the nematic $\mathbf{d}_o$-vector represents a $(s+d)$-wave pairing states,
\begin{align}
\begin{cases}
\text{s-wave dominant: } , t_{1,2} \ll 1, \text{fully gapped superconductors} \\ 
\text{d-wave dominant: } , t_{1,2}\gg 1, \text{ndoal superconductors}
\end{cases}
\end{align}
As a result, we have $|\mathbf{d}_o|\sim|\mathbf{g}_{tot}|=|1+k_F(\theta)\left({t_1}\sin2\theta+t_2
\cos2\theta\right)|$. We see that the SC gap function can have nodes as long as $\sqrt{{t_1^2}+t_2^2}$ is large enough. For graphene, the FS has $C_3$ symmetry, i.e. $k_F(\theta)$ has periodicity of $\frac{2}{3}\pi$, whereas $\sin2\theta$ and $\cos2\theta$ have periodicity of $\pi$, giving a periodicity of $2\pi$ to $|\mathbf{d}_o|$, which completely breaks the $C_3$ symmetry of the system. Figure~\ref{sm-fig2} shows the $C_3$-breaking nematic orders from inter-valley scattering, one with nodal gap function, the other with nodeless gap function.
\begin{figure}[!htbp]
	\centering
	\includegraphics[width=0.35\linewidth]{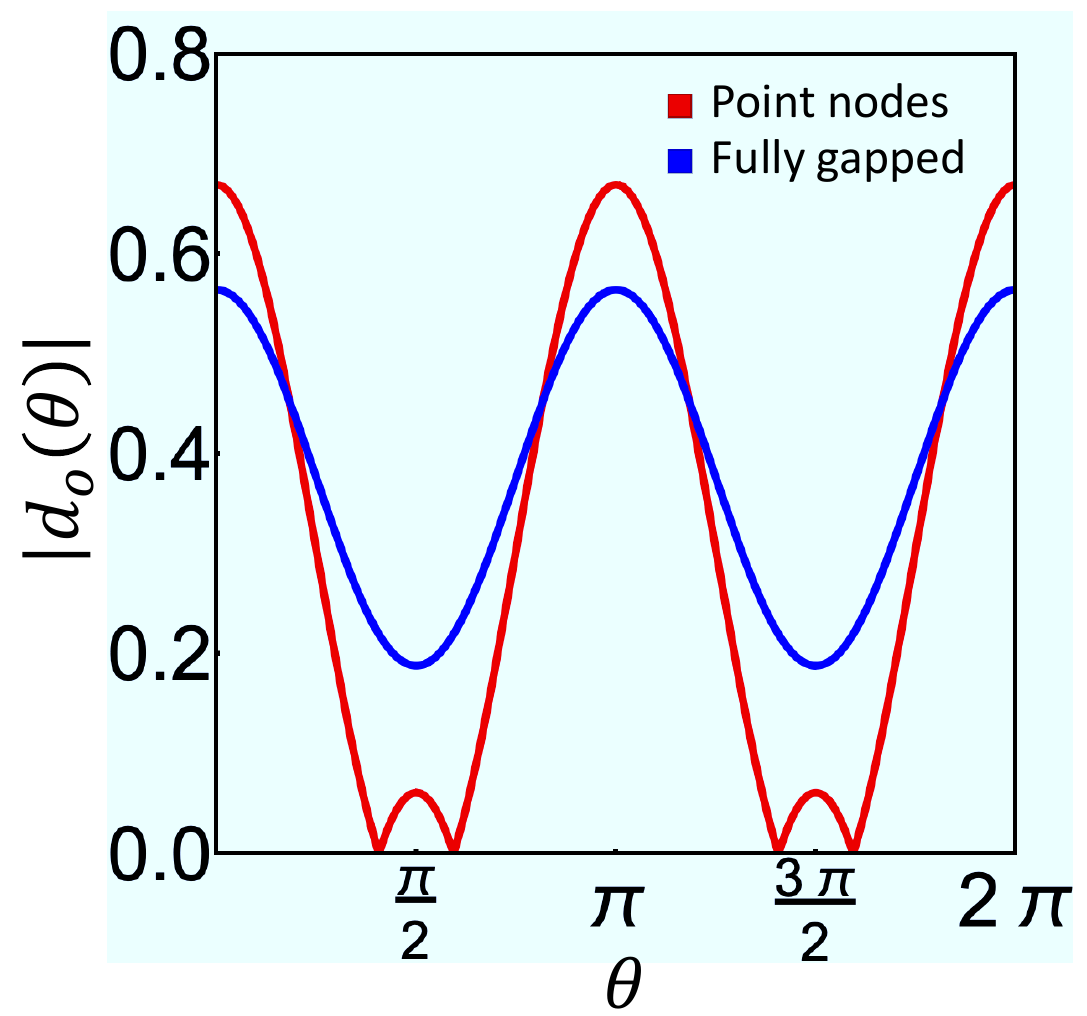}
	\caption{The $C_3$-breaking nematic order from inter-valley scattering in graphene. $t_1=0,t_2=1.2$ for the nodal case (red) and $t_1=0,t_2=0.5$ for the fully gapped case (blue).
	}
	\label{sm-fig2}
\end{figure}

\section{Spin and orbital magnetizations: $\mathbf{M}_s$ and $\mathbf{M}_o$}
Here we present the definitions of spin and orbital magnetizations in our mean-field analysis.
The spin magnetization is defined as,
\begin{align}
\mathbf{M}_s= \sum_{\mathbf{k},s_1,s_2}\langle c_{s_1}^{\dagger}(\mathbf{k})\bm{\sigma}_{s_1s_2}c_{s_2}(\mathbf{k})\rangle,
\end{align}
and the orbital magnetization is defined as,
\begin{align}
\mathbf{M}_o= \sum_{\mathbf{k},s,a,b}\langle c_{s,a}^{\dagger}(\mathbf{k})\bm{\tau}_{ab}c_{s,b}(\mathbf{k})\rangle.
\end{align} 
More specifically, the orbital magnetization has the following components, 
\begin{align}
M_o^x&=\sum_{\mathbf{k},s}\langle c_{s,d_{xz}}^{\dagger}c_{s,d_{yz}}+c_{s,d_{yz}}^{\dagger}c_{s,d_{xz}}\rangle, \\ M_o^y&=-i\sum_{\mathbf{k},s}\langle c_{s,d_{xz}}^{\dagger}c_{s,d_{yz}}-c_{s,d_{yz}}^{\dagger}c_{s,d_{xz}}\rangle=\frac{1}{2}\sum_{\mathbf{k},s}\langle \hat{n}_{s,d_{xz}+id_{yz}}-\hat{n}_{s,d_{xz}-id_{yz}}\rangle, \\
M_o^z&=\sum_{\mathbf{k},s}\langle c_{s,d_{xz}}^{\dagger}c_{s,d_{xz}}-c_{s,d_{yz}}^{\dagger}c_{s,d_{yz}}\rangle, 
\end{align}
where $M_o^{x,z}$ breaks the $C_4$ rotation symmetry and $M_o^y$ breaks the TRS.
In our work, we focus on spontaneous TRS breaking, with $C_4$ preserved.
$M_o^y(\mathbf{k})$ gives the local density difference in mementum space for $d_{xz}\pm id_{yz}$ orbitals. A non-zero $M_o^y(\mathbf{k})$ implies TRS breaking at $\mathbf{k}$, because the two orbitals are TR partners. However, it has to be noted that, similar to the spin-triplet case, local TRS breaking does not necessarily imply the global TRS is broken. To obtain the total overall magnetization, we still need to sum over momentum around the FS. In the Ginzburg-Landau formalism, the $\mathrm{Im}[\mathbf{d}_o\times \mathbf{d}_o^*]$, whose only non-zero component is the $y$-component based on symmetry constraints in our formalism,  is coupled to the induced magnetization $M_o^y$ so that TRS is still retained at the Lagrangian level. Therefore, we have the following relevant terms
$$\alpha M_o^y\mathrm{Im}[\left(\mathbf{d}_o\times \mathbf{d}_o^*\right)_y]+\beta \left(M_o^y\right)^2,$$
which upon functional derivative with respect to the induced magnetization would give
$$M_o^y\propto \mathrm{Im}[\left(\mathbf{d}_o\times \mathbf{d}_o^*\right)_y]=-i\left(\mathbf{d}_o\times \mathbf{d}_o^*\right)_y.$$

\end{widetext}
\end{document}